\pdfoutput=1
\documentclass[structabstract]{aa}
\usepackage{graphicx}
\usepackage{txfonts}
\usepackage{array}
\usepackage{natbib}
\bibpunct{(}{)}{;}{a}{}{,}
\usepackage{longtable}
\usepackage{float}
\usepackage{color}
\usepackage{dcolumn}
\newcolumntype{k}[1]{D{.}{.}{#1}}
\newcolumntype{d}{D{.}{.}{-1}}
\newcolumntype{,}{D{,}{,}{-1}}

\begin{document}
\title{IRC +10 216 in 3-D: morphology of a TP-AGB star envelope.}

\author{
 M. Gu\'elin\inst{1,2}, N.A. Patel\inst{3}, M. Bremer\inst{1}, J. Cernicharo\inst{4}, A. Castro-Carrizo\inst{1}, J. Pety\inst{1}, J.P. Fonfr\'ia\inst{4},
M. Ag\'undez\inst{4},   M.~Santander-Garc\'ia\inst{4}, G. Quintana-Lacaci\inst{4}, L. Velilla Prieto\inst{4}, R. Blundell\inst{3}, P. Thaddeus\inst{3}}

\institute{Institut de Radioastronomie Millim\'etrique, 300 rue de la Piscine, 38406 Saint Martin d'H\`eres, France; guelin@iram.fr 
\and 
LERMA, Observatoire de Paris, PSL Research University, CNRS, UMR 8112, F-75014, Paris, France
\and 
Center for Astrophysics, 60 Garden street, Cambridge, MA USA
\and 
ICMM. CSIC. Group of Molecular Astrophysics. C/ Sor Juana In\'es de la Cruz 3. Cantoblanco, E-28049 Madrid, Spain}

\titlerunning{IRC +10 216 in 3-D}
\authorrunning{Gu\'elin et al.}

\abstract{
{During their late pulsating phase, AGB stars expel most of their
mass in the form of massive dusty envelopes, an event that largely
controls the composition of interstellar matter. The envelopes, however,
are distant and opaque to visible and NIR radiation: their structure
remains poorly known and the mass-loss process poorly
understood. Millimeter-wave interferometry, which combines the
advantages of longer wavelength, high angular resolution and very high
spectral resolution is the optimal investigative tool for this
purpose.  Mm waves pass through dust with almost no attenuation. Their
spectrum is rich in molecular lines and hosts the fundamental lines of
the ubiquitous CO molecule, allowing a tomographic
reconstruction of the envelope structure.}
{The circumstellar envelope IRC +10 216 and its central star, the C-rich TP-AGB star closest to the Sun, are the best objects for such an
investigation. Two years ago, we reported the first detailed study of the
CO(2-1) line emission in that envelope, made with the IRAM 30-m
telescope. It revealed a series of dense gas shells, expanding at a
uniform radial velocity. The limited resolution of the
telescope (HPBW $11\arcsec$) did not allow us to resolve the shell
structure. We now report much higher angular resolution observations of
CO(2-1), CO(1-0), CN(2-1) and C$_4$H(24-23) made with the SMA, PdB and
ALMA interferometers (with synthesized half-power beamwidths of
$3\arcsec$, $1\arcsec$ and $0.3\arcsec$, respectively).} 
{Although the envelope appears much more intricate at high resolution
than with an $11\arcsec$ beam, its prevailing structure remains a
pattern of thin, nearly concentric shells. The average separation between the
brightest CO shells is $16\arcsec$ in the outer envelope, where it
appears remarkably constant. Closer to the star ($<40\arcsec$), the
shell pattern is denser and less regular, showing intermediary arcs.}
{Outside the small ($r<0.3''$) dust formation zone, the gas appears to
expand radially at a constant velocity, 14.5 km\,s$^{-1}$, with small
turbulent motions.  Based on that property, we have reconstructed the
3-D structure of the outer envelope and have derived the gas
temperature and density radial profiles in the inner ($r<25''$) envelope. The shell-intershell density
contrast is found to be typically 3. The over-dense shells have spherical or
slightly oblate shapes and typically extend over a few
steradians, implying isotropic mass loss.} {The regular spacing of shells
in the outer envelope supports the model of a binary star system with
a period of 700 years and a near face-on elliptical orbit. The
companion fly-by triggers enhanced episodes of mass loss near
periastron. The densification of the shell pattern observed in the central
part of the envelope suggests a more complex scenario for the last few thousand years.}
 \thanks{This work was based on observations carried out with the IRAM,
SMA and ALMA telescopes.  IRAM is supported by INSU/CNRS (France), MPG
(Germany) and IGN (Spain).The Submillimeter Array is a joint project
between the Smithsonian Astrophysical Observatory (USA) and the Academia
Sinica Institute of Astronomy and Astrophysics (Taiwan) and is funded by the
Smithsonian Institution and the Academia Sinica. This paper makes use
of the ALMA data: ADS/JAO.ALMA\#2013.1.01215.S \& ADS/JAO.ALMA\#2013.1.00432.S. 
ALMA is a partnership
of ESO (representing its member states), NSF (USA) and NINS (Japan),
together with NRC (Canada), NSC and ASIAA (Taiwan), and KASI (Republic
of Korea), in cooperation with the Republic of Chile. The Joint ALMA
Observatory is operated by ESO, AUI/NRAO and NAOJ.}
}
\date{Received July 24,2017; accepted September 10, 2017}
\maketitle
\keywords{astrochemistry -- stars: AGB and post-AGB -- circumstellar matter --
stars: individual (IRC +10 216)}

\section{Introduction}

Three-quarters of the matter returned to the interstellar medium (ISM)
comes from AGB stars in their late thermally pulsating (TP)
phase. The mass loss rate may then reach 10$^{-5}$ to 10$^{-4}$ $M_{\sun}\,\cdot $yr$^{-1}$
and the stars become enshrouded by a thick, dusty envelope opaque
to visible and near IR radiation. Given that the TP phase is short-lived,
the closest AGB-TP stars are fairly distant. High visual opacity and 
distance conspire to make the mass loss mechanism and the envelope structure 
poorly understood, despite their importance for galactic evolution.

The advent of powerful millimeter/sub-mm interferometers that combine
longer wavelengths with high angular and spectral resolution provides an
unique opportunity to investigate these objects. Mm/sub-mm waves
freely traverse the thickest dust layers, and the mm spectrum, rich in
molecular lines, yields detailed information on the gas physical
conditions, its chemical content and, mostly, on the velocity
field. The latter often reduces to uniform radial expansion, allowing
the 3-dimensional envelope structure to be recovered. We present in
this paper the first high resolution study of the entire IRC +10 216
envelope, the archetype of TP-AGB star envelopes, carried out in the J=2-1 line 
of CO, the best single tracer of the molecular gas in this type of object
\citep{Ramstedt2008}.

The envelope IRC +10 216\footnote{Designation from the two-micron sky survey 
(\citet{Neugebauer1969}), composed of the survey declination 
strip and a serial number. The name IRC $+10^\circ$216, was 
condensed into IRC +10 216 by \citet{Becklin1969}.} 
(CW~Leonis), which surrounds the closest C-rich
TP-AGB star to the Sun (hereafter denoted {\it CW~Leo$\star$}), is of particular
interest. Located at a distance of $\sim 130$~pc \citep{Menten2012},
it appears on optical images as a dark spherical cloud that extends
over several arcmin. Some 80 molecular species, half of all known
interstellar molecules, have been detected in this envelope through
thousands of millimeter-wave lines \citep{Cernicharo2000}. Remarkably,
all line profiles but those arising from the tiny dust-acceleration
region have the same width in the direction of the star: 29
kms$^{-1}$. Since the molecular lines arise at different distances
from the central star -- e.g. SiO and SiS are restricted to the
central region, whereas the CO is distributed over the whole envelope
-- the constant line width implies a steady expansion velocity along
the line of sight, 14.5 kms$^{-1}$, and a small turbulent
velocity. The systemic velocity of the envelope is -26.5 kms$^{-1}$
relative to the local standard of rest (LSR).



IRC +10 216 has been the object of numerous studies in so far as 
dust and molecular content are concerned. Of particular relevance to this work are
{\it a)} the CFHT, Hubble and VLT V-light images of the scattered IS light
(\citet{Mauron2000,Mauron2006} and \citet{Leao2006}), which revealed
the presence of a ringed dust structure (Fig.~\ref{COonFORS1}); {\it b)} the detection of UV emission
from a termination shock (or {\it astrosphere}) that marks the 
impact of the outflowing gas on the surrounding ISM at $15'$ ($\simeq 2$pc) NE 
of CW~Leo$\star$ \citep{Sahai2010}; and {\it c)} the FIR emission map made with the PACS
instrument on Herschel \citep{Decin2011}. The 100 $\mu$m map, like
the visible light images, shows a succession of rings that mark the rims
of dense spherical shells. Finally, {\it d)} interferometric images of the mm
line emission of a score of reactive molecules, such as CN, CCH and
HC$_5$N, show that these species are mostly confined inside a thin
spherical shell of radius 15$''$ whose center, curiously, is offset by
2$''$ from CW~Leo$\star$ \citep{Guelin1993b}.

\begin{figure}
\includegraphics[angle=0,width=0.8\columnwidth]{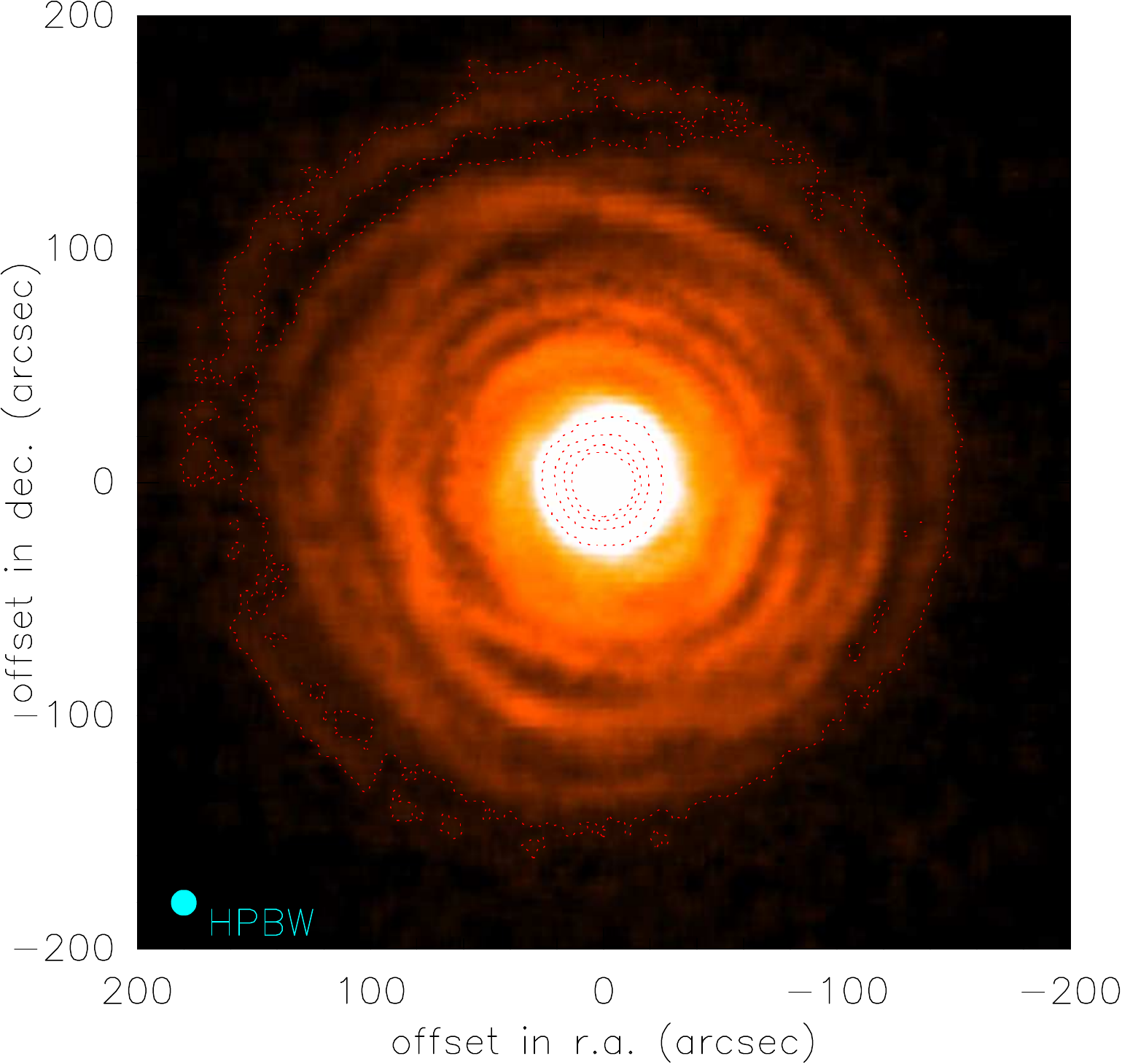}   
\caption{\label{SDv-26map} Main beam-averaged $^{12}$CO(2-1) line brightness temperature observed in IRC +10 216
with the IRAM 30-m telescope (HPBW 11$''$) at the star velocity (V$_*=-26.5$ km\,s$^{-1}$, 
$\Delta$v=2 km\,s$^{-1}$ --see \citet{Cernicharo2015}. Dotted contours range from 1 to 50 K. On 
this, as on the other velocity-channel maps, the brightness scale is linear.}
\end{figure}
\begin{figure}
\includegraphics[angle=0,width=0.32\columnwidth, trim=30 0 0 0,clip=true]{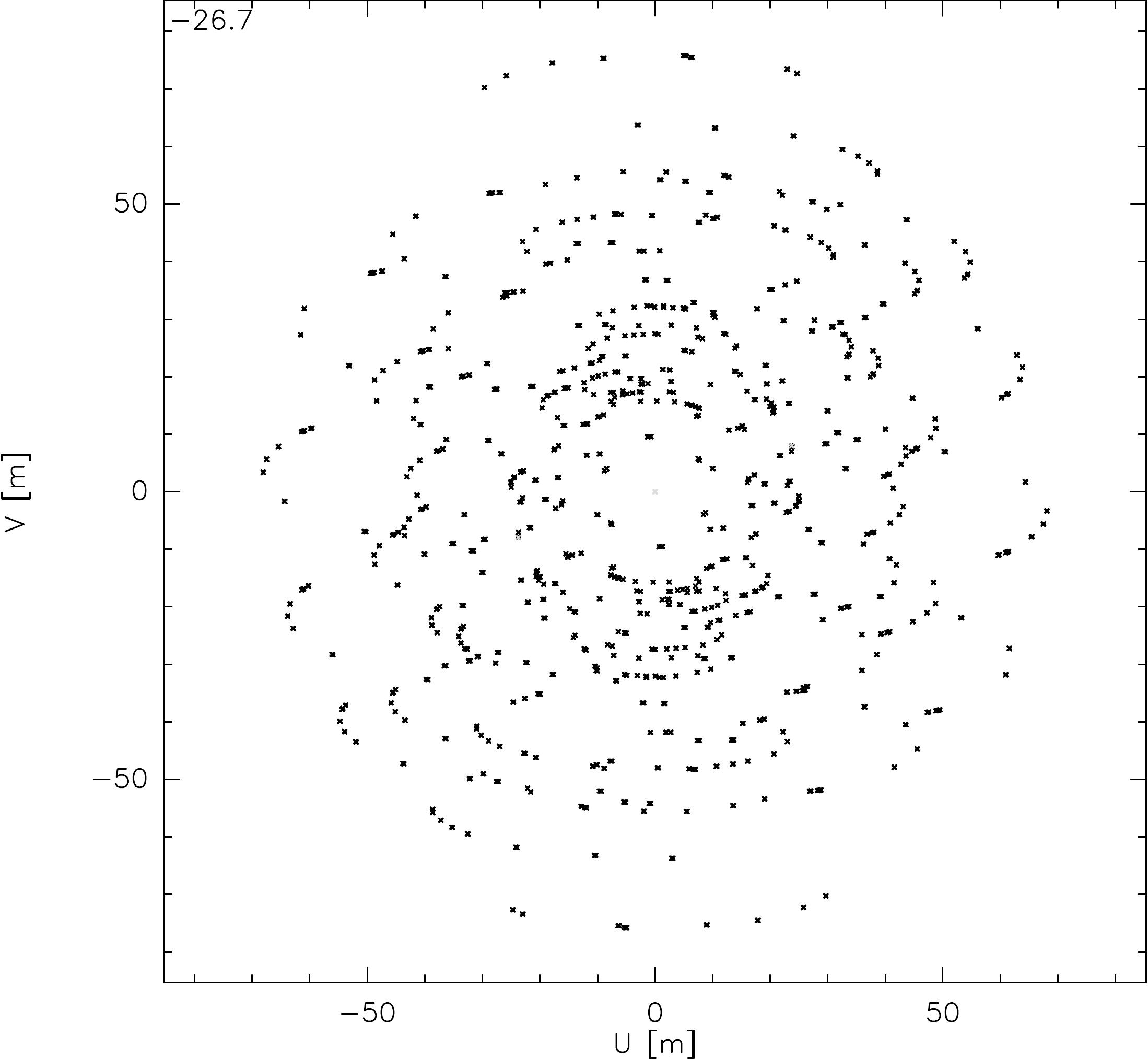}
\includegraphics[angle=0,width=0.32\columnwidth,trim=30 0 0 0,clip=true]{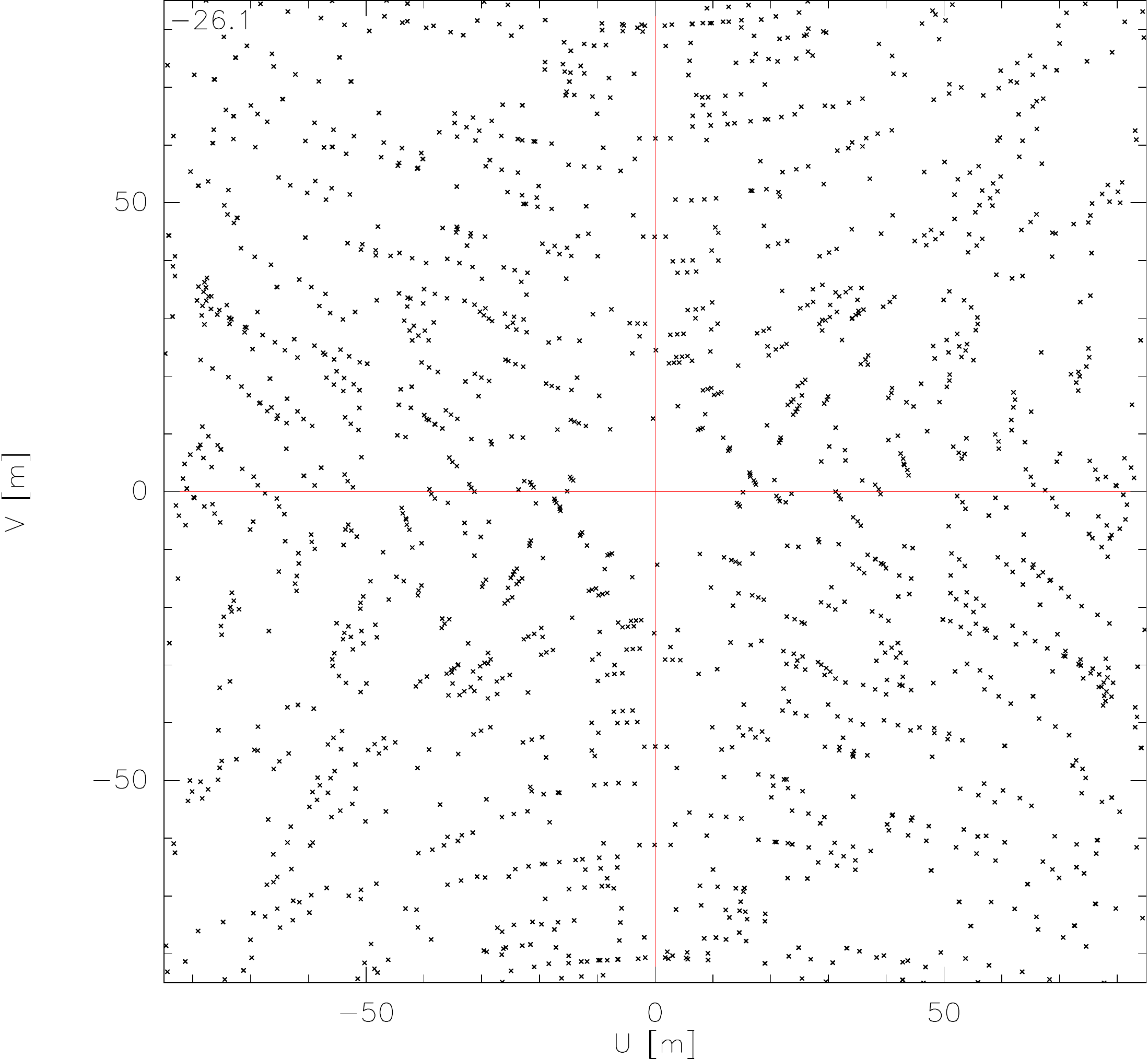}
\includegraphics[angle=0,width=0.32\columnwidth]{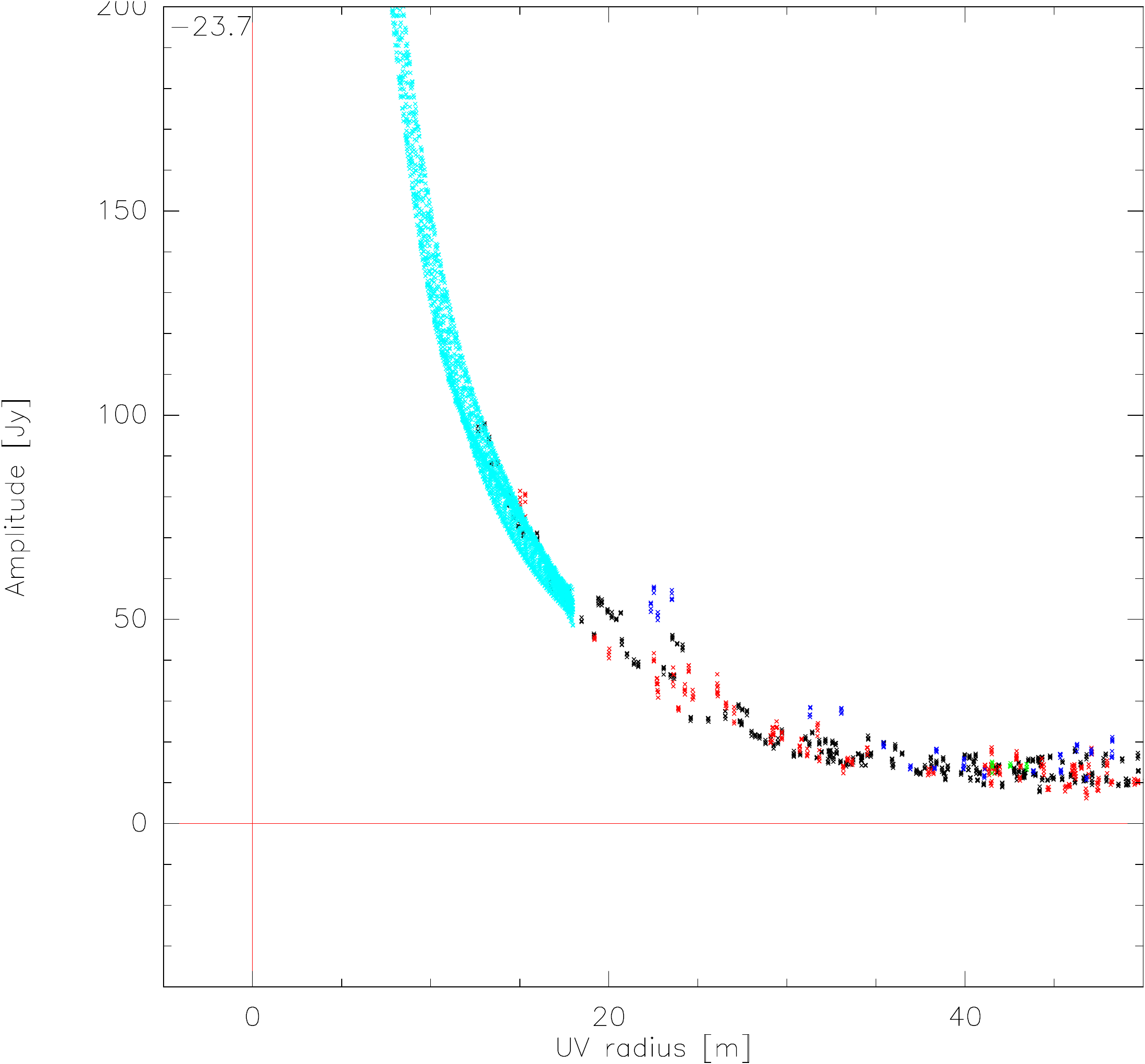}
\caption{\label{uvcov} Observed visibilities in the central 100 m of the UV-plane for the SMA {\it left} 
and ALMA {\it middle} observations.
Thanks to the subcompact configuration, the shortest spacings for the SMA are as small as 6-m. {\it Right} 
Visibility amplitudes versus radius for the merged ALMA+single dish observations. The 30-m telescope amplitudes 
(blue dots) have been scaled by a factor of 0.8 to match the ALMA ones (red dots) in the overlap region.}
\end{figure}

We have previously reported a detailed study of IRC +10 216, made in the
$^{12}$CO(2-1) and $^{12}$CO(1-0) lines (hereafter denoted CO(2-1) and CO(1-0)) 
and in the $^{13}$CO(1-0) line  with the IRAM 30-m telescope (Cernicharo,
Marcelino, Ag\'undez and Gu\'elin 2015). The CO(2-1) line emission was mapped throughout
the entire envelope at a resolution of 11$''$ (HPBW) and the line was
continuously detected up to the photodissociation radius $r_{phot}\simeq 180''$, 
inside which CO efficiently self-shields from
interstellar UV radiation. The CO envelope fits well inside the large 
bow-shock traced by the UV emission, which suggests it expands freely
inside the cavity cleared up by the shock (both the $^{12}$CO and the
$^{13}$CO line profiles have the canonical full width of 29 kms$^{-1}$).
It consists of a bright central peak, a broad, slowly decreasing pedestal 
and, superimposed on the pedestal,  a series of over-dense gas shells. 

\begin{center}
\begin{figure*}[th!]
\includegraphics[angle=0,width=0.94\textwidth]{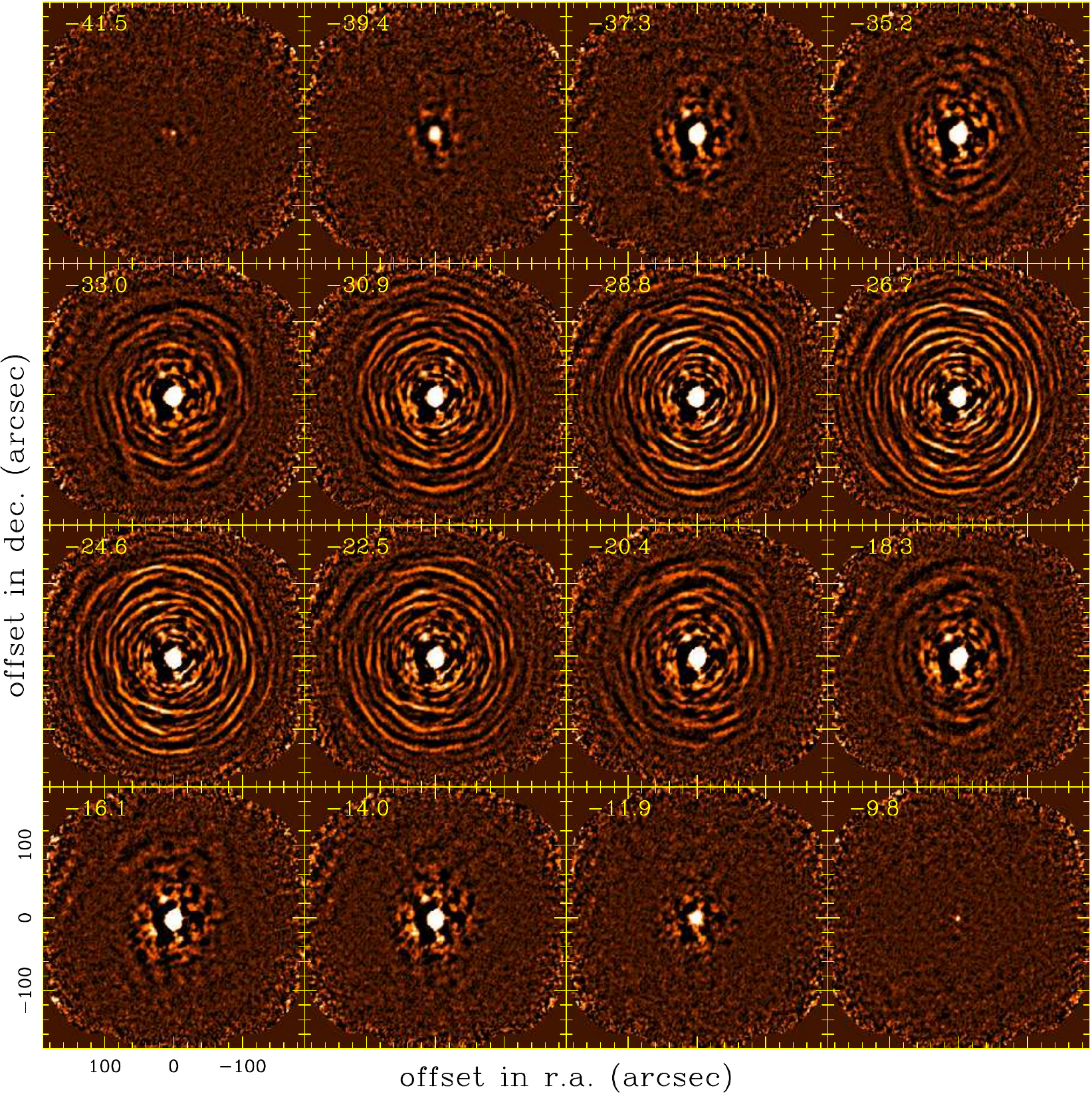}
\caption{
\label{SMAvcmaps}
Velocity-channel maps of the CO(2-1) line emission in the central $400''\times 400''$ area 
of the IRC +10 216 envelope, observed with the SMA at a spatial resolution of $3.7''\times 3.0''$ (HPBW, PA=-51). The spectra have been binned to a resolution of 2.1 km\,s$^{-1}$);
velocities are relative to the LSR; the star systemic velocity is $V_*=-26.5\,$km s$^{-1}$; 
velocities more positive or more negative correspond to the rear and front parts of the envelope, respectively. The lack of visibilities $<6$ m carves a negative bowl and black areas at the center of the maps (see text).
}
\end{figure*}
\end{center}

The limited angular resolution of the 30-m telescope did not allow us
to resolve the shell structure. So, we turned to the SMA, PdB and ALMA
interferometers for much higher resolution observations. The envelope
was fully mapped in CO(2-1) with the SMA at $3\arcsec$ resolution, and
partly with PdBI at $1\arcsec$ resolution. The central 1-arcmin region
was mapped in the CO(1-0), $^{13}$CO(1-0), CN(2-1) and C$_4$H(24-23)
lines with the ALMA 12-m antennas and, for the short spacings, with
the IRAM 30 m-diameter single-dish (SD) telescope. In Sec. 2, we describe the observations and
data reduction procedures, and in Sec. 3 \& 4 derive the envelope
morphology, velocity field and gas physical properties. Finally,in Sec. 5, we
interpret the results in terms of mass-loss by a binary
star system.

\section{Observations and data reduction}

\begin{center}
\begin{figure*}
\includegraphics[angle=0,width=0.945\textwidth]{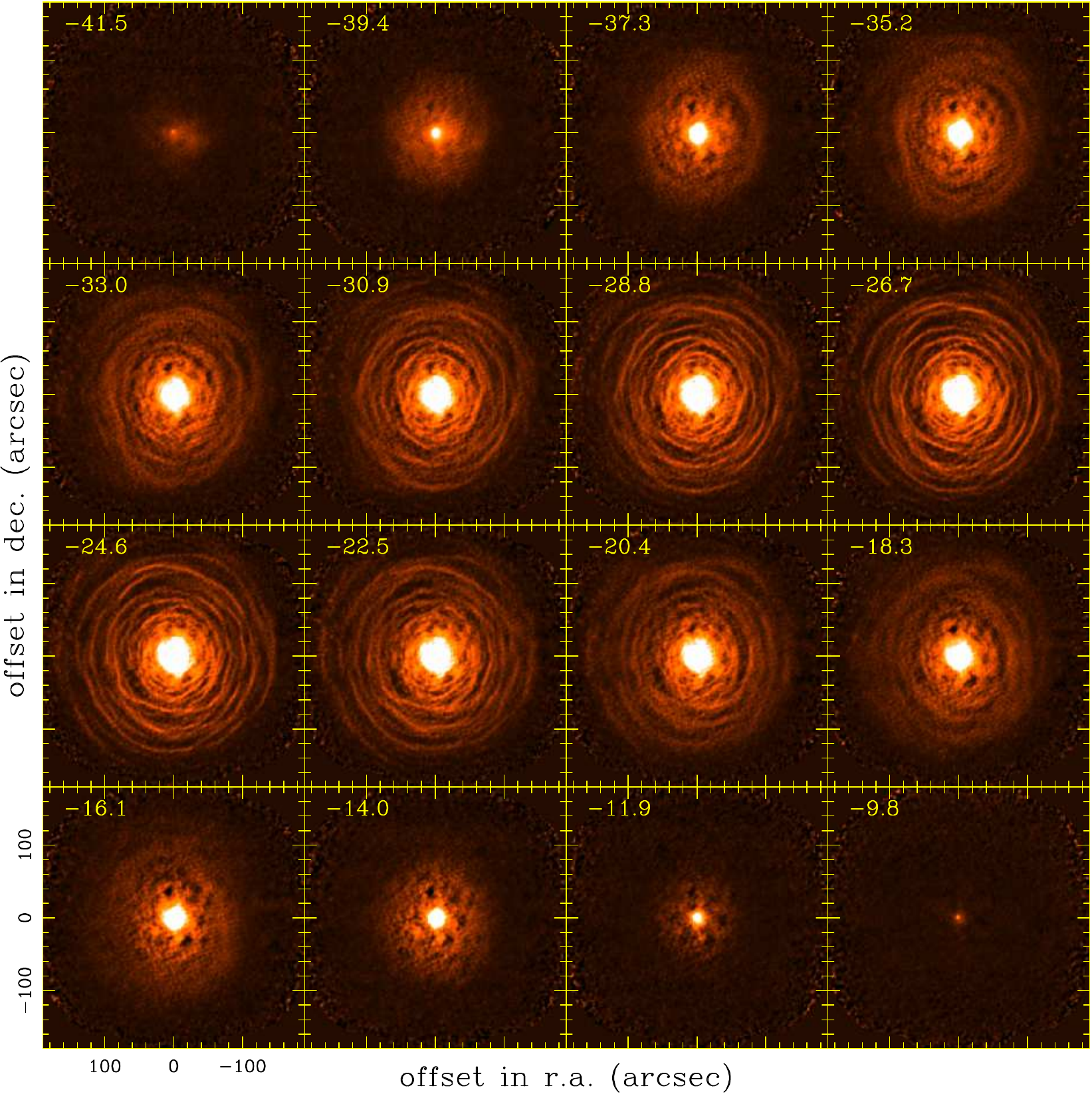}
\caption{
\label{SMASDvcmaps}
Velocity-channel maps of the CO(2-1) line emission after combining the SMA data and 
the 30-m SD telescope data (see text). The spatial resolution is $3.8''\times 3.1''$ (HPBW PA=-51$^\circ$). The spectra 
have been binned to a resolution of 2.1 km\,s$^{-1}$); marked velocities are expressed in km\,s$^{-1}$
and relative to the LSR. The 
star velocity is $V_*=-26.5\,$km s$^{-1}$.
}
\end{figure*}
\end{center}

\begin{center}
\begin{figure*}[t]
\includegraphics[angle=0,width=0.945\textwidth]{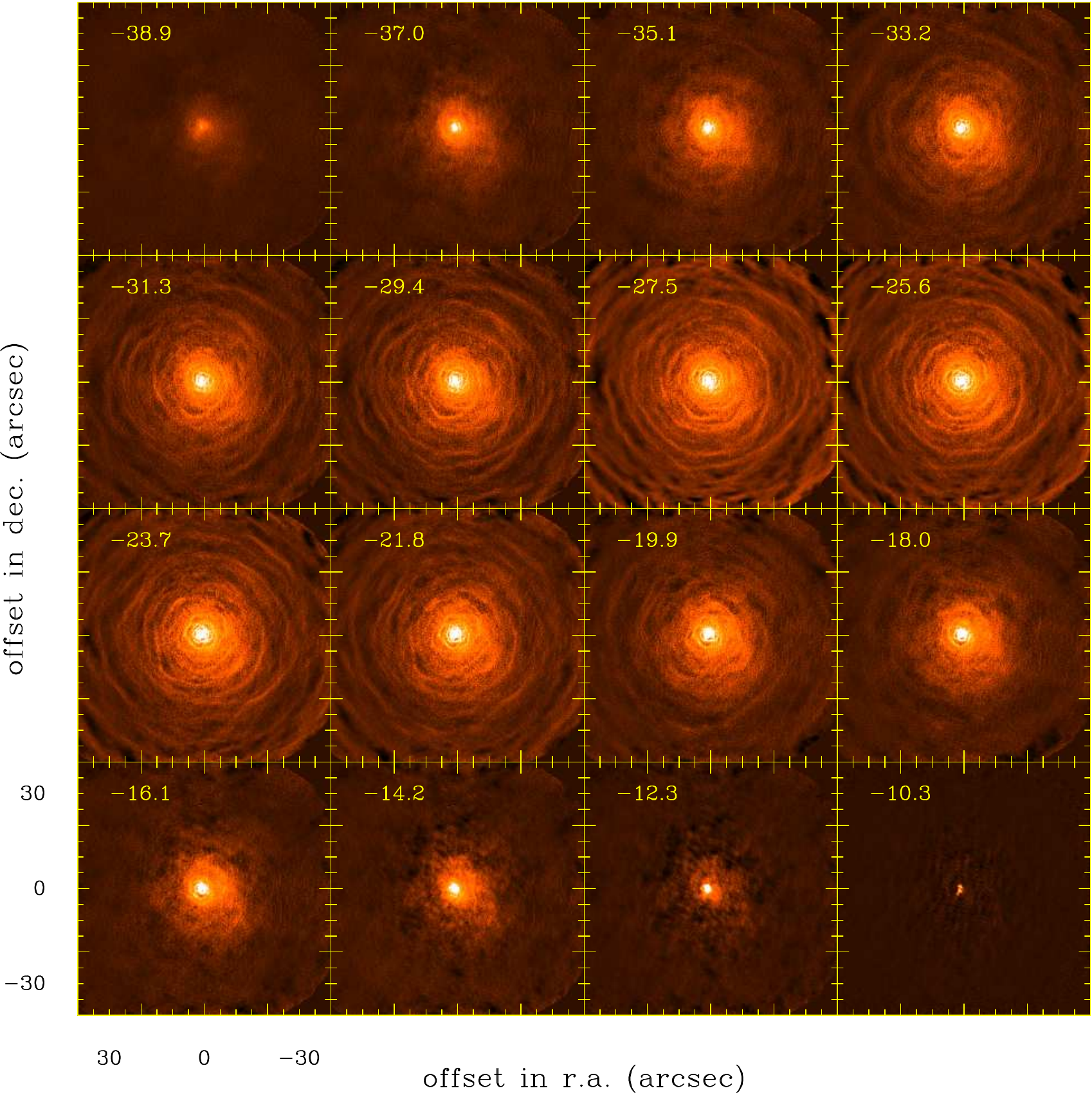}
\caption{
\label{ALMAvcmermaps}
Velocity-channel maps of the CO(2-1) line emission in the central $80''\times 80''$ area 
of the IRC +10 216 envelope, observed with ALMA+SD at a spatial resolution of $0.35''\times 0.33''$ (HPBW PA=38$^\circ$). The spectra 
have been binned to a resolution of 1.9 km\,s$^{-1}$; marked velocities are expressed in km\,s$^{-1}$
and relative to the LSR. The 
star velocity is $V_*=-26.5\,$km s$^{-1}$; 
}
\end{figure*}
\end{center}

\subsection{SMA observations}
The SMA observations were made between Dec 2011 and May 2013 in the
Compact configuration and in May 2013 in the Subcompact
configuration. Depending on the observing period, the array consisted
of 7 or 8 x 6-m diameter antennas. The antenna HPBW at the frequency
of the CO(2-1) line (230.53797 GHz) is 54.6$''$. The observations were
made during Winter and Spring, mostly during night time. The
precipitable water vapor was $< 4$ mm and the sky zenith opacity at
220 GHz between 0.1 and 0.3. The receivers were tuned so that the
CO(J=2-1) line was in the LSB, centred on unit s16 of the 4 GHz
bandwidth {\it ASIC} spectral correlator. The spacing between adjacent frequency
channels was originally 812 kHz, but the data in Fig.~\ref{SMAvcmaps} were binned to 1625 kHz
({2.1} kms$^{-1}$ at the CO(2-1) line frequency) to reduce the channel noise and
suppress the overlap between adjacent channels.

We used the mosaic mode through a series of
observing loops of typically 50 pointings, spaced by 25$''$ (0.45
HPBW) in RA and DEC. Two such loops were observed during a typical
10 h track. A total of 157 pointings were observed
covering a nearly circular area of diameter $\simeq 400''$. The signal
phase and amplitude were calibrated by observing the quasars 0854+2006
and 0927+390 every 20 minutes. The antenna pointing was checked prior
to each observing loop and the flux scale calibrated by observing
Titan and/or Callisto on every track, as is standard in SMA
observations. The receiver bandpass was also calibrated by observing
3C279 on every track.

The SMA data were calibrated using the {\it MIRIAD} calibration
package \citep{Sault1995}. Calibrated {\it uvfits} files were converted into {\it GILDAS} format and
further data processing and analysis was made with the
{\it GILDAS/MAPPING/MOSAIC} data reduction software \footnote{http://www.iram.fr/IRAMFR/GILDAS}. The continuum signal
was derived for each pointing by averaging the line emission-free
channels of the 4 GHz-wide correlator and was subtracted from the original
data to yield the CO(2-1) line data.  The latter were first processed
alone to produce a first position-velocity datacube, then reprocessed
after completing the SMA visibilities with short spacing
pseudo-visibilities from the 30-m SD telescope,
to produce a second position-velocity datacube. The steps in the
second reduction were: {\it a)} processing the SD data to derive
pseudo-visibilities compatible with the SMA visibilities of each mosaic field; {\it b)} checking
the relative calibration of both instruments; {\it c)} adding for each field the
single-dish pseudo-visibilities with spatial frequencies between 0 and
20 m to the SMA visibilities; {\it d)} Fourier transform the so-merged
visibilities to derive the CO(2-1) {\it dirty} map of each field; {\it e)} correct these maps 
for primary beam attenuation after truncating them at the 20\% attenuation level and combine 
them into a single dirty image; finally {\it f)}
deconvolve the dirty map by the dirty beam to produce the final clean images, using a modified Hogb\"om
algorithm that proceeds iteratively by order of decreasing signal-to-noise ratio. 

\begin{center}
\begin{figure*}
\includegraphics[angle=0,width=0.945\textwidth]{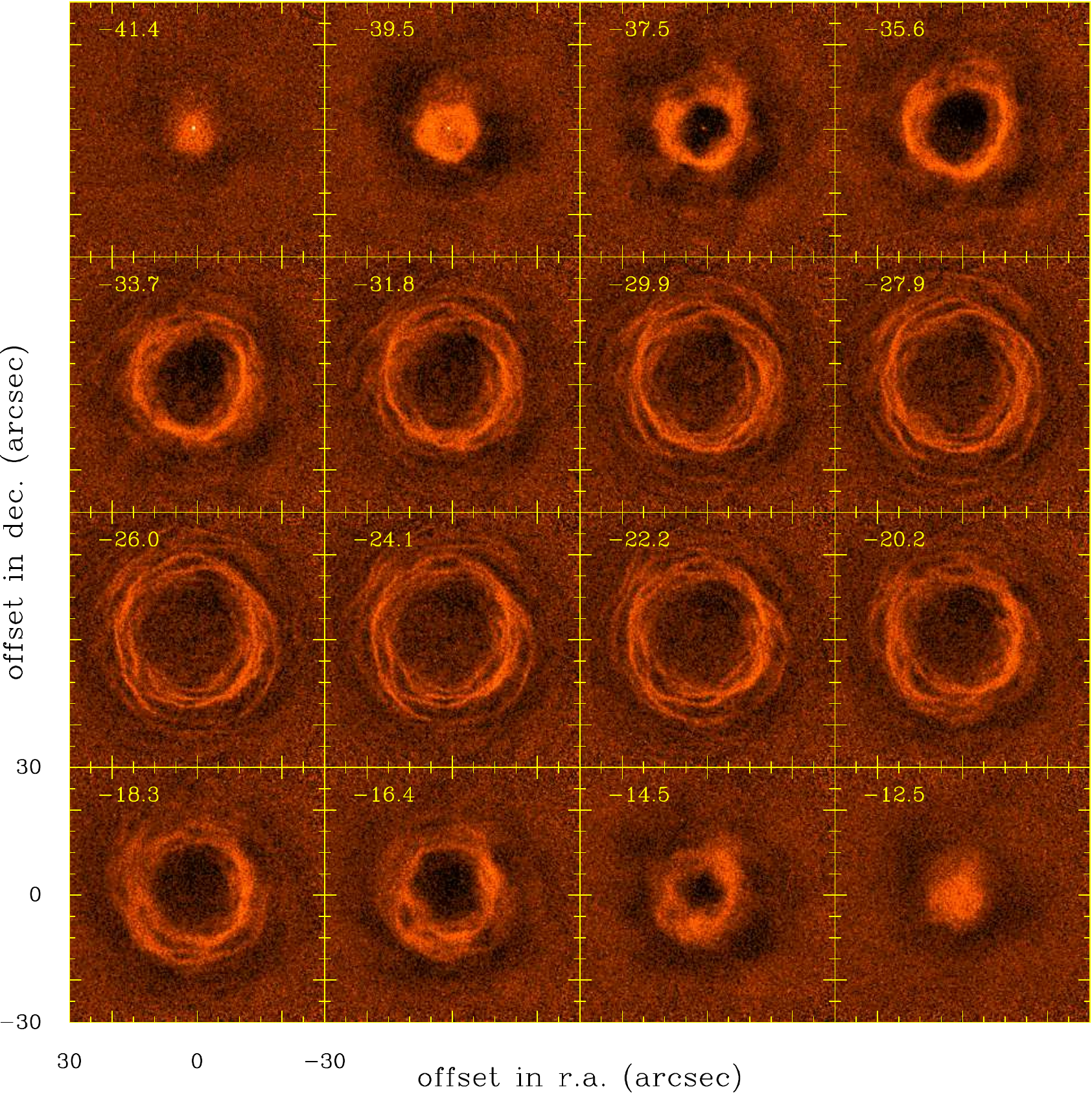}
\caption{
\label{ALMAvcC4Hmaps}
Velocity-channel maps of the C$_4$H(N=24-23) line emission in the central $60''\times 60''$ area 
of the IRC +10 216 envelope, observed with ALMA. 
The spectra have been binned to a resolution of 1.9 km\,s$^{-1}$; marked velocities are relative to the LSR. The 
star velocity is $V_*=-26.5\,$km s$^{-1}$; 
velocities more positive or more negative correspond to the rear and front parts of the envelope, respectively.
}
\end{figure*}
\end{center}

The whole (a) to (f) procedure is as described in the {\it GILDAS}
{\it MAPPING} Documentation.  Step (b) was made by comparing the amplitudes
of the azimuthally averaged pseudo-visibilities of both instruments
in the uv-plane region where they overlap. Thanks to the Subcompact
configuration, the SMA visibilities extended down to the physical size
of the antennas, providing a comfortable range of overlap: from 6 m to
24 m. This allowed us in step (c) to discard the single-dish
pseudo-visibilities with radii larger than 20 m, which critically
depend on the beam pointing and shape, and are therefore less
reliable. The amplitudes of the single-dish pseudo-visibilities were scaled
by a factor 1.7 to match the azimuthally-averaged SMA amplitudes in
the overlap region. No apodisation was applied during step (e),
i.e. the SMA visibilities were processed with their natural weights;
the weights of the single-dish pseudo-visibilities, on the other hand,
were lowered to a constant value to avoid an undue widening of the
synthesized beam. 

The uv-plane coverage with the SMA and with ALMA is illustrated in Figure~\ref{uvcov}.
The mosaicing observing mode allows, in principle, for recovery of part of the visibilities for spacings
$<6$ m. However, since these have a low weight, the reconstruction of the SMA images remains
problematic, particularly near the extended central source. 
    
The final SMA and SMA+SD image datacubes are shown in Figures~\ref{SMAvcmaps}
\& \ref{SMASDvcmaps} in the form of velocity-channel maps of velocity resolution 2.1
kms$^{-1}$. The angular resolution (synthesized beam HPW) is $3.7''
\times 3.0''$ (PA=-51$^\circ$) for the SMA maps and $3.8'' \times 3.1''$ (PA=-51$^\circ$) for
the SMA+SD maps. The r.m.s. noise in the maps varies from 0.25
mJy\,beam$^{-1}$, in the region within $\pm 50''$ in DEC and $\pm 120''$ in RA 
from CW~Leo$\star$, to 0.35
mJy\,beam$^{-1}$ at the map edges. The CO line emission shows a very
strong source at the position of CW~Leo$\star$. This source is so
bright that it causes strong negative lobes in the dirty map at radii
$r \leq 30''$, lobes that turned out impossible to clean despite the 
mosaicing observing mode, in view of
the limited dynamic range of the SMA data. To obtain reliable images
of this region, we have re-observed the central 1 arcmin with ALMA.

\subsection{ALMA observations}
The 1-mm ALMA observations (project code 2013.1.01215.S) were made with the 12-m array in the Compact
configuration (Dec 2014 and Jan 2015, 32 antennas) and in the Extended
configuration (June 2015, 39 antennas). Ganymede and the quasars 0854+2006
and J1008+0621 were observed on every track for flux and phase calibration. 
The baselines ranged from 
12.7 m to 1569.4 m. The observing mode consisted of a 27-field
mosaic with adjacent pointings spaced by 11.2$''$ in RA and 12.9$''$
in DEC, i.e. spaced by 0.42 and 0.48 times the 12-m antenna primary
beam (HPBW 26.8$''$ at the CO(2-1) line frequency). The mosaic covered 
a square area of $60'' \times 60''$ at
full sensitivity, extending to $90'' \times 80''$ at $1/5^{th}$ sensitivity.  Besides
CO(2-1), the usable band of the ALMA Band 6 receiver covered several
molecular lines of interest, the most conspicuous being the three upper
CN(N=2-1) fine structure (fs) components (around 226.9 GHz), both C$_4$H (N= 24-23) 
fs components (228.3486 and 228.3870 GHz) and, in the receiver upper-sideband, the CS(J=5-4) 
line (244.9350 GHz). Those were simultaneously observed with CO  using several 
spectral correlator windows with different spectral resolutions. The channel 
separation for the CO,
CN, C$_4$H and CS lines was 0.2 kms$^{-1}$, 0.3 kms$^{-1}$ 0.4
kms$^{-1}$ and 0.6 kms$^{-1}$, respectively; most of the data
presented here were binned to resolutions of 1 or 2 kms$^{-1}$.

One subband of the correlator was tuned to fully cover 
the 244.2-246.1 GHz sky frequency interval from the receiver upper sideband. 
The only truly strong line in this interval is the 244.935 GHz
line of CS(5-4). After discarding the channels with CS emission, we used 
this subband to derive the continuum emission.
   
The data were calibrated through the ALMA/{\it CASA} pipeline. The
resulting {\it uvfits} files were converted into the {\it GILDAS uvt} format and
further processing was made with the {\it GILDAS/MAPPING/MOSAIC}
software package. The process was similar to that just described for the SMA
data: two CO(2-1) image datacubes were produced, one for ALMA data,
the other for the merged ALMA plus SD data. The channel-velocity 
maps, binned to velocity resolutions of 1.9 kms$^{-1}$ 
are shown on Figure~\ref{ALMAvcmermaps} for ALMA+SD CO and on Fig.~\ref{ALMAvcC4Hmaps}
for ALMA C$_4$H. The synthesized HPBW is $0.34''
\times 0.31''$ (PA=+36$^\circ$) for the ALMA maps and $0.35'' \times 0.33''$ (PA=+38$^\circ$) for the
ALMA+SD maps. The r.m.s noise is 3 mJy/beam per 2 kms$^{-1}$ channel and  
50 $\mu$Jy/beam in the continuum 1.9 GHz subband. A comparison of the 
ALMA+SD map, smoothed to the resolution of the SMA+SD map, shows the same 
bright structures than the latter, outside the innermost region ($r>20''$).

The 3-mm ALMA observations (project code 2013.1.00432.S) were obtained in 2015 with the 12-m array in the Compact 
and Extended configurations. They consisted in a single pointing, centered on the position
of the star in 2012.4 (see Table 1), and were part of a spectral survey covering the 3-mm 
atmospheric window. 
The field of view (half-power primary beam width) of the 12-m ALMA antennas at 115.5 GHz is 54$''$. 
The 3-mm data were processed like the 1-mm data and similarly combined with 30-m SD data.
The synthesized HPBW was $0.48'' \times 0.45''$ for the CO(1-0) line. A thorough
description of the spectral survey will be presented elsewhere
(Cernicharo et al., {\it in preparation}).


\subsection{IRAM 30-m single-dish telescope and PdBI observations}
The CO(2-1) 30-m SD observations were reported in \citet{Cernicharo2015}.  
They mainly consisted of a fully sampled
square map of size 480$'' \times$
480$''$, plus 4 additional 240$'' \times$ 240$''$ maps centred at the
corners of this square.  All maps were observed on-the-fly and a 
low order baseline was subtracted from each of the dumped spectra. 
The baseline subtraction removed the receiver instabilities, but also 
the continuum signal. The
telescope radiation pattern mainly consists of a Gaussian main beam,
of half-power width (HPBW) 11$''$, plus a much wider error beam
containing 25\% of the total power. The contribution of the error beam
was subtracted in the maps and analysis presented here. Fig.~\ref{SDv-26map}
shows the emission seen by the 30-m SD telescope at the velocity of the star CW~Leo$\star$.
These observations have been used as short spacings for the SMA and ALMA  
CO $J$=2-1 line data presented here.

Additional CO(1-0) and $^{13}$CO(1-0) observations, consisting of a fully sampled map of 
size $120''\times 120''$, centred on the star, were made in the Fall 2015 with the 30-m telescope, as 
part of a spectral survey \citep{Agundez2017,Quintana2017}.

Finally, fully sampled $120''\times 120''$ maps of the  C$_4$H(24-23), CN(2-1) and 
CS(5-4) line emissions were carried out in December 2016 with the 30m IRAM telescope.
They provide the total flux and short spacings for the ALMA maps. 

Prior to the SMA and ALMA observations, we made preliminary CO(2-1)
interferometric observations with the PdBI 6 antenna 
array. They consisted in a mosaic covering one quarter of a 60$''$-wide annulus of inner 
radius 50$''$. The synthesized HPBW was $1.4\arcsec \times 1.2 \arcsec$ 
and the noise 30 mJy/beam in 3 kms$^{-1}$-wide channels. Since the same region
was re-observed with the SMA with a higher sensitivity and denser uv plane coverage, the 
PdBI data will be used only in Sec. 3.1, when we compare the inner to the outer arc 
pattern.

\begin{figure*}[!ht]
\centering
\includegraphics[angle=0,width=0.9\textwidth]{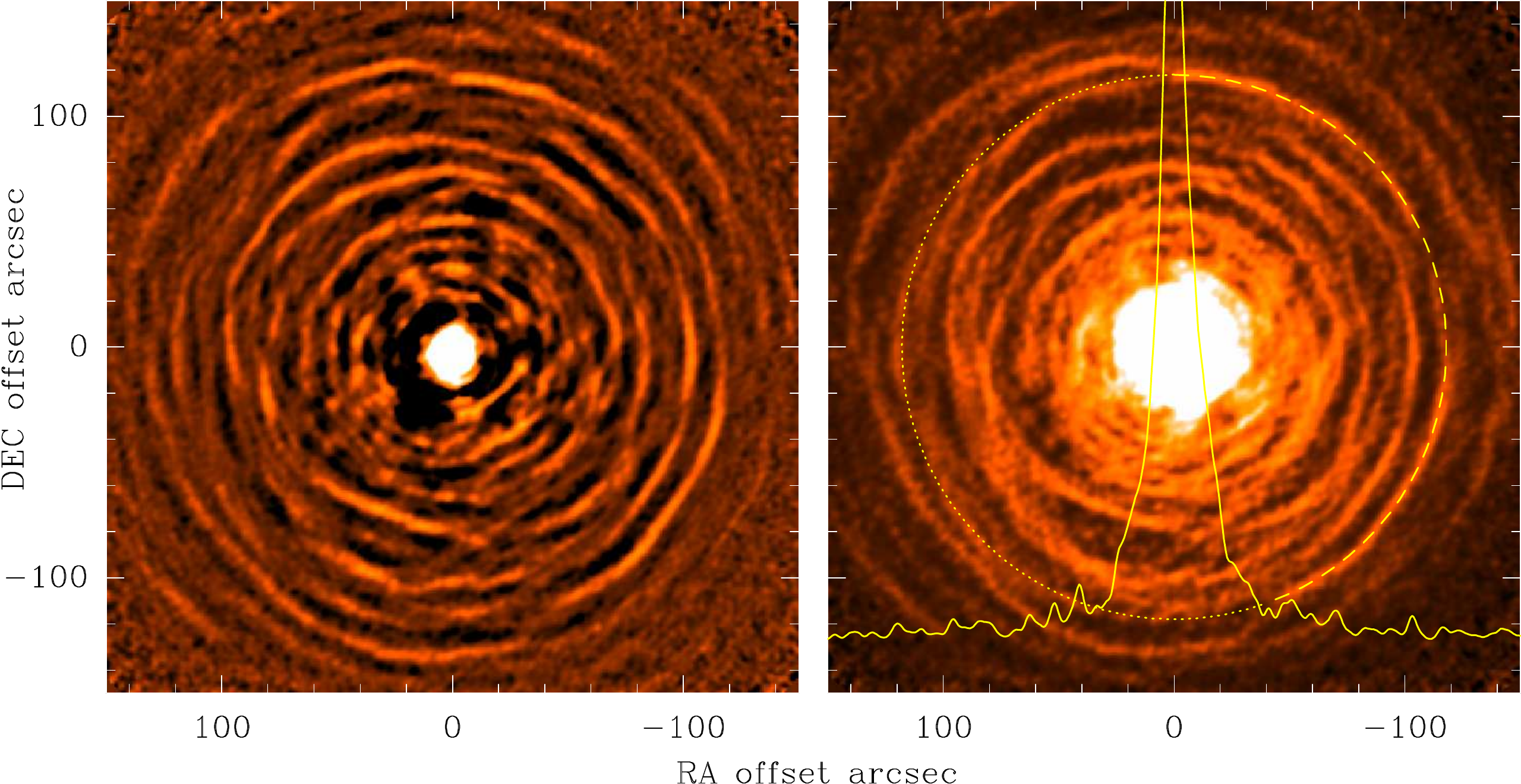}
\caption{\label{SMAv-26map} Maps of the CO(2-1) line emission at the star velocity (-26.5 kms$^{-1}$) viewed by the SMA 
before ({\it left}) and after ({\it right} inclusion of the 30-m single-dish short spacings. The yellow curve 
shows the intensity profile along an EW strip passing through the central star. Most of the bright CO arcs 
have circular shapes and denote the intersection of the meridional plane with thin spherical shells. The 
latter typically extend over several steradians, as illustrated by the arc at      
$r=118''$ in the western half of the envelope (dashed yellow circle). The dense gas shell 
associated with this arc can be followed over $\pi$ radians along the line of sight on 
Fig.~\ref{EWplane-it}.}
\end{figure*}

\section{Results}

\subsection{Gas distribution in the meridional plane}
\subsubsection{Outer envelope}

As pointed out above, the outer envelope appears to be expanding with a
constant velocity of 14.5 km\,s$^{-1}$ and has a small ($< 1$ km\,s$^{-1}$) turbulence velocity,
judging from the remarkably constant width of the molecular lines (we
further discuss this point below in Sec. 4.1, when dealing with the 3-D
structure of the envelope). Each velocity-channel map (of e.g. Figs.~\ref{SMASDvcmaps} and ~\ref{ALMAvcmermaps})
shows the emission from a conical sector (of opening $\theta$ and
thickness $\delta \theta$), with its axis aligned to the line of
sight to the star CW~Leo$\star$ (see Fig.~\ref{fig-reco-1} in Appendix A). The extreme velocities 
-40 km\,s$^{-1}$ and
-12 km\,s$^{-1}$ correspond to the approaching and receding polar
cones and caps, while the central velocity $V_*$= -26.5
km\,s$^{-1}$ corresponds to a cut through the envelope at the star
position; the CO envelope being roughly spherical, the $V_*$= -26.5
km\,s$^{-1}$ map appears more extended.

Figure~\ref{SMAv-26map} shows an enlargement of the CO $V= -26.7$
km\,s$^{-1}$ map before ({\it left panel}) and after ({\it right panel}) addition of the SD
short spacings. Thanks to the filtering
of the extended emission component by the interferometer, the left map
shows more clearly the bright ringed structure noted in previous
observations. The bright arcs, which trace dense shells of molecular
gas in the outer envelope (see Cernicharo et al. 2015), appear far
more clearly at the 3$''$ resolution of the SMA than at the 11$''$
resolution of the 30-m SD (Figure~\ref{SDv-26map}). Surprisingly, they look
almost circular and seem regularly spaced, except for three 
nearly straight segments in the SE quadrant. In contrast, the pattern
in the inner 30$''$ radius region appears confused in Figure~\ref{SMAv-26map},
due to artefacts in the image restoration process. We discuss the central region below.

\begin{figure}[!hb]
\includegraphics[angle=0,width=0.7\columnwidth]{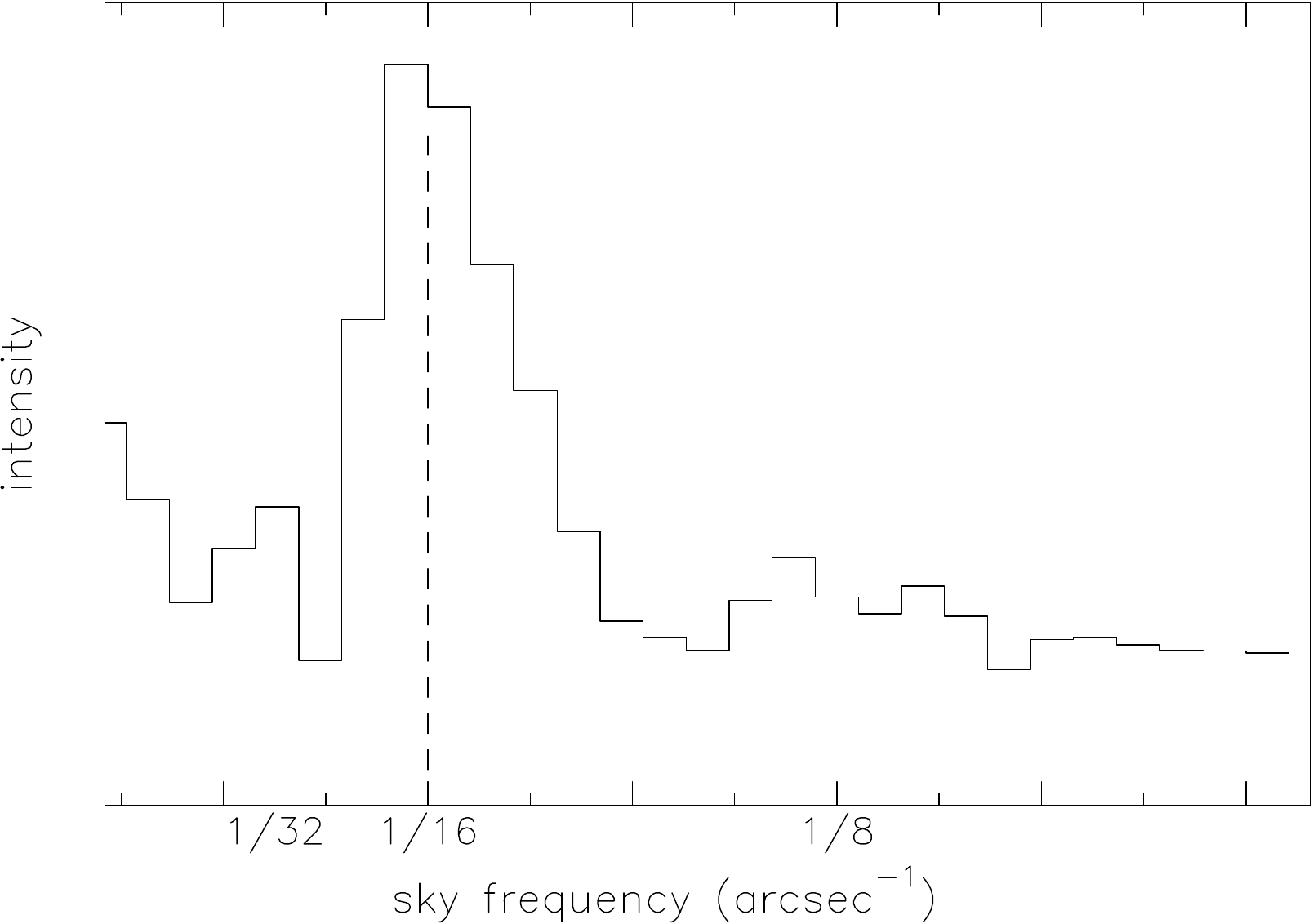}
\caption{\label{fft} Fourier transform of the azimuthally averaged CO(2-1) line brightness distribution for the 
W half of the envelope. Abcissa is spatial frequency in arcsecond$^{-1}$.
}
\end{figure}

We have analyzed the ring pattern in the outer envelope by azimuthally
averaging the SMA map of Fig.~\ref{SMAv-26map} in the NW, NE, SE and
SW quarters and by Fourier transforming the resulting 4 radial
profiles from $r= 20''$ to $r=150''$. Except for the SE quadrant, the
FT shows a marked peak that corresponds to a period of 16$''$
(Fig.~\ref{fft}), i.e. to a time lag of 700 yr for an expansion
velocity of 14.5 km\,s$^{-1}$ and a distance of 130 pc. Most of bright
rings of the outer envelope can be fairly well fitted in each quadrant
by circular arcs. 

The circular arcs are centred close to CW~Leo$\star$, albeit not exactly on. 
This was first 
noted by \citet{Guelin1993b}, who studied the shape of the inner
bright ring at $r\simeq 15''$ appearing on the C$_4$H maps and found 
that it was centred a couple of arcsec NE of the star. They
interpreted the offset as a drift of the dense shell, caused by an orbital motion 
of CW~Leo$\star$. The brightest arcs
in Fig.~\ref{SMAv-26map} may be similarly fitted by circular arcs slightly offset; the 
offsets range between 2$''$ and 10$''$. 

\begin{figure}[hb!]
\includegraphics[angle=0,width=0.95\columnwidth]{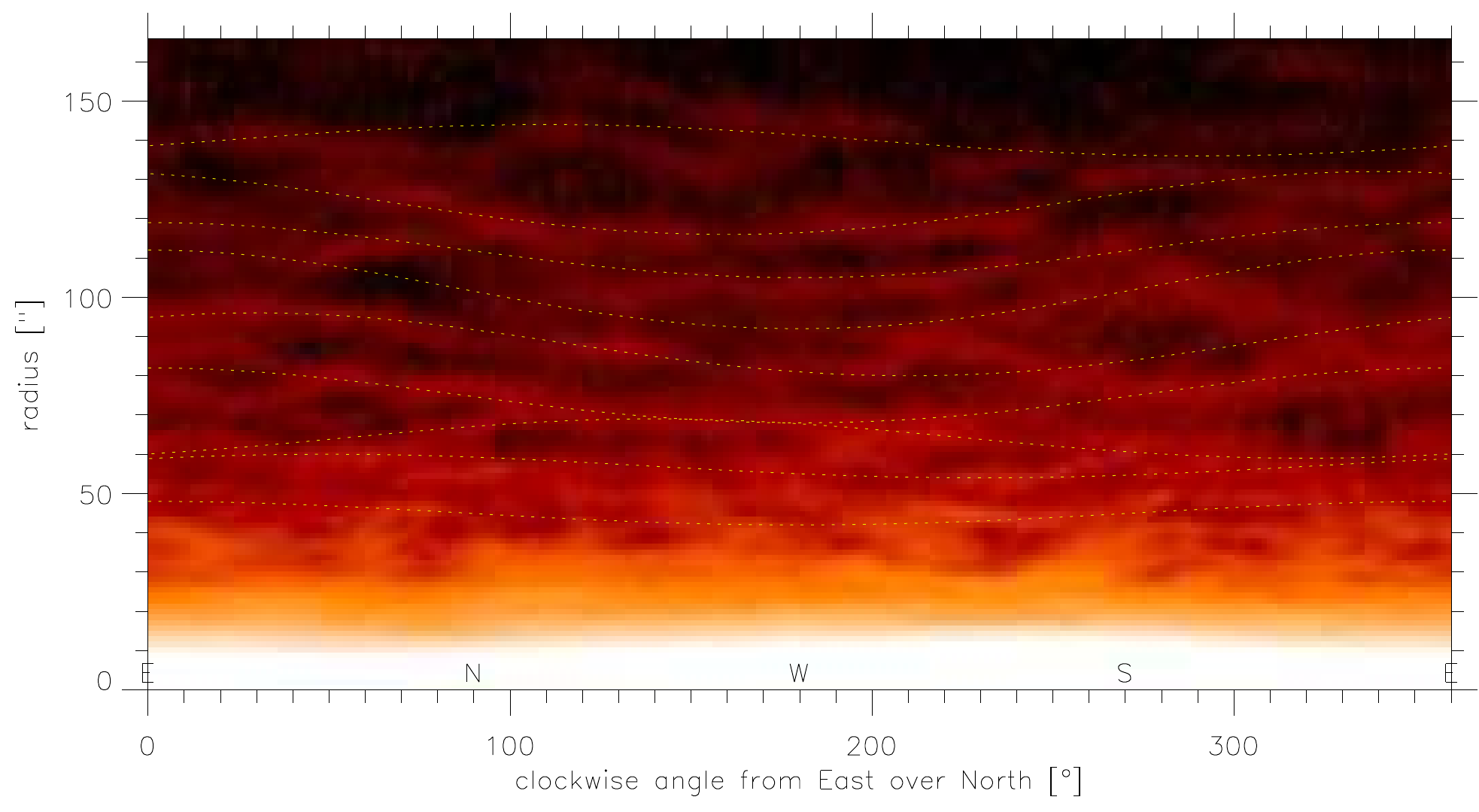} 
\caption{\label{SMApolarview} 
The CO(2-1) brightness distribution of Fig.~\ref{SMAv-26map} in polar coordinates and logarithmic scale. The 
yellow sinusoids traced over the brightest features correspond to slightly off-centred circles in the 
meridional plane. 
}
\end{figure}

We note that the fits become poorer when replacing
the circular arcs by segments of spirals with significant pitch angles.  
This is illustrated in Fig.~\ref{SMApolarview},
which shows the CO $V= -26.7$ km\,s$^{-1}$ emission in polar
coordinates. In that representation, an Archimedean spiral would appear as a tilted 
straight line. We observe instead 100-150$^\circ$-long segments almost
parallel the x-axis, bending up or down near their edges by up to 10$''$, i.e. the signature of 
slightly off-centred circular arcs. The bending of several bright segments (e.g. at $r=100''$ and $130''$)
occur around angles of $170^\circ\pm 180^\circ$, implying an off-centering along 
the E-W axis. Globally, the offset appears to increase with radius. 

\subsubsection{Central region}
\begin{center}
\begin{figure*}[!ht]
\includegraphics[angle=0,width=0.95\textwidth]{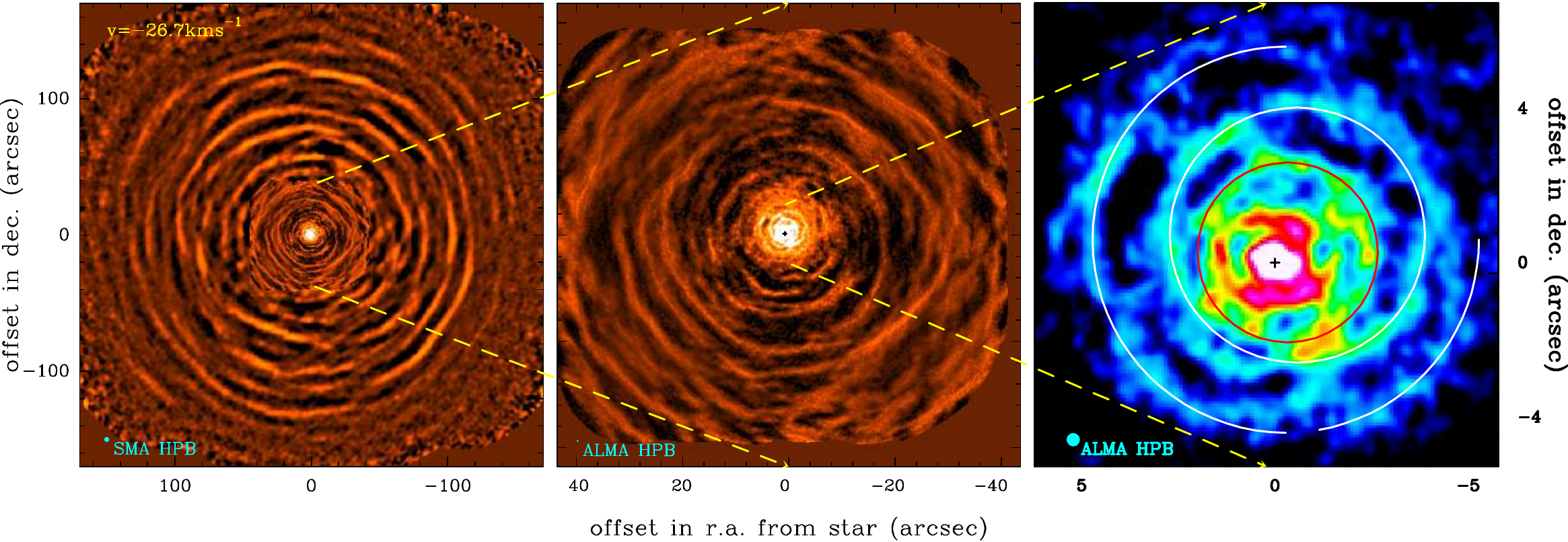}
\caption{
\label{SMAinsert}
The CO(2-1) emission at the star velocity ($V_*=-26.5$ km\,s$^{-1}$) at different scales: 
{\it left frame} $380''\times 380''$ SMA map (HPBW 3$''$) with, at its center, the 
$90'' \times 80''$ 27 field mosaic ALMA map; {\it center frame} ALMA map (HPBW $0.34''\times 0.31''$)
at an enlarged scale; and {\it right frame} zoom on the $12'' \times 12''$ inner region 
of the envelope. No SD data has been added. The color code in the right panel is chosen to underline
the bright arcs. The circles illustrate a possible fit of the arc-like features with 
slightly off-centred circular arcs. 
The blue dots show the SMA and ALMA half-power beam widths. Position (0,0 black cross) corresponds to the star
 09$^h$\,47$^m$\,57.4492$^s$ 13$^\circ\,16'\,43.884''$ --see Table~1.}
\end{figure*}
\end{center} 

The CO emission at the star velocity in the central part of the envelope is shown at 
different scales in the 3 panels of Fig~\ref{SMAinsert}: {\it left frame} $380''\times 380''$ SMA map
with at its center the $90'' \times 80''$ 27 field mosaic ALMA map ; {\it center frame} the ALMA map 
at an enlarged scale; {\it right frame} a zoom into the central $12'' \times 12''$ region of 
the ALMA map. No SD data has been added in this figure.

The remarkable periodicity of the bright arcs in the outer envelope appears to break down
within an arcmin of the star. This is not just an effect of the tenfold increase 
in angular resolution and sensitivity. Indeed, closer than 50$''$ from the star, the SMA map
shows the same contracted arc pattern as the ALMA data, smoothed to the $3''$ SMA resolution; 
similarly the arc pattern observed between 50$''$ and 110$''$ E of CW~Leo$\star$ with PdBI, smoothed to 3$''$, 
is the same as that in (Figure~\ref{SMAv-26map}). 

Between $r=10''$ and $r=40''$, the spacing between the brightest arcs is $\simeq 10''$ 
in the NW quadrant and further decreases to $\simeq 2''$ closer to the star. The spacing is 
even smaller in the opposite (SE) quadrant, possibly the effect of a NW drift of 
the arc centers. We note that the innermost arcs in  
Fig~\ref{SMAinsert} seem to form a series of off-centred rings, rather 
than a single regular spiral. This is 
illustrated on the right frame by the fit to the inner arc pattern of
3  circles with radii 2.3$''$, 3.3$''$ and 5$''$, respectively centred at 
(0.35$''$,0.45$''$), 
(-0.6$''$,0.3$''$) and 
(-0.8$''$,0.5$''$) from the star.  
We note that the arc centres seem to drift to the W with increasing radius, i.e. 
with the arc age. The arcs are fully resolved with the 0.3$''$ ALMA beam. They appear as tangled filaments,
some of which are kinked or show bridging branches.


The appearance of the dense shells in the CO(2-1) line, particularly on 
Fig.~\ref{SMASDvcmaps}, is confused by the bright CO source around the star and 
by the CO emission in the intershell region. In order to have a clearer view 
of the inner shells, we have turned to two radicals, namely C$_4$H
and CN, known to form far from the star. The C$_4$H(N=24-23) and CN(N=2-1) line frequencies are
close enough to that of the CO(2-1) line, and were observed simultaneously with the ALMA receiver. 

Figure~\ref{ALMAvcC4Hmaps} shows the velocity-channel maps of the upper frequency 
component of the  C$_4$H (N=24-23) line doublet, observed with ALMA. The continuum emission
has been removed. As expected, no emission is detected near the star, except for foreground
and background emission at terminal velocities (near -12 km\,s$^{-1}$ and -40 km\,s$^{-1}$).
We also note a point-like source near -38 km\,s$^{-1}$
of width $\simeq 6$ km\,s$^{-1}$ characteristic of vibrationally excited lines, 
that can be assigned to Si$^{34}$S $\nu$=1 J=13-12 at $\nu$=228399.382 MHz.
Lines from SiS and its isotopologues involving vibrationally excited states 
from $\nu$=1 up to $\nu$=10 have been
detected with ALMA by \citet{Cernicharo2013} 
and in the infrared by \citet{Fonfria2015}.

It was previously noted by \citet{Guelin1993b} that the C$_4$H emission is essentially confined 
to a thick shell extending from $r=10''$ to $r=25''$; the brightest contours at the star velocity 
($V_*=-26.5$ km\,s$^{-1}$) trace a thick circular ring centred $2''$ NE of the star.  
With the 0.3$''$ ALMA beam, each shell is resolved into $\simeq 1''$-thick  
arcs with different radii and different centers of curvature. The outermost arc to the SW, which 
on the -26 km\,s$^{-1}$ map may be interpreted as the onset of a spiral, rather appears 
as a circle 
centred at (0.6$''$,-2.6$''$) on the adjacent -24 km\,s$^{-1}$
and -22 km\,s$^{-1}$ velocity-channel maps (see Fig.~\ref{ALMAvcC4Hmaps}).

\begin{center}
\begin{figure}
\includegraphics[angle=0,width=0.45\textwidth]{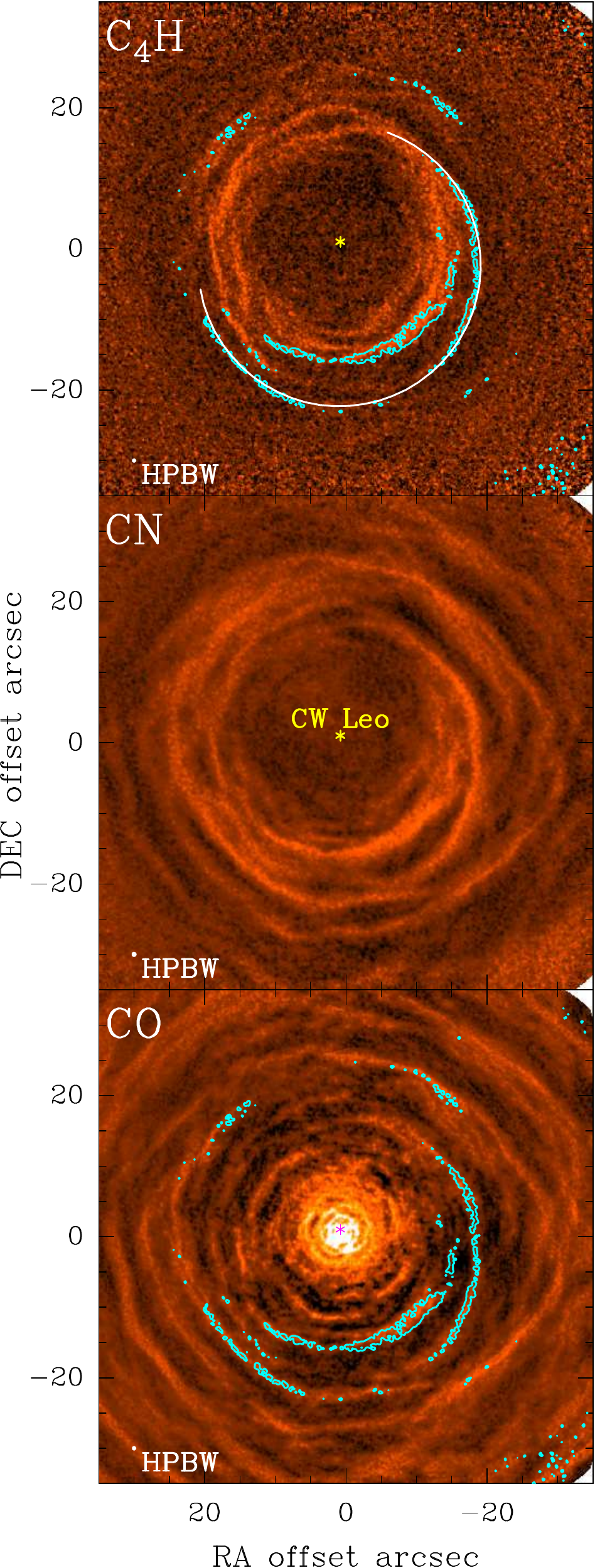}
\caption{
\label{COCNC4H} The C$_4$H(24-23) and CN(2-1) emissions at the star velocity ($V_* = -26.5$ km\,s$^{-1}$), observed with ALMA 
(HPBW $0.34''\times 0.31''$), compared to the CO(2-1) emission. The cyan color contours correspond to 
the 0.25 Jy/beam level from the CN map. The star indicates the position of CW~Leo$\star$. 
The white circular arc on the upper frame is a fit to the bright southern 
arc; it is centred 0.6$''$ E,2.6$''$ S of CW~Leo$\star$. 
}
\end{figure}
\end{center}

This arc structure is not specific to C$_4$H, but also found for the
CN(2-1) line emission. Figure~\ref{COCNC4H} shows the C$_4$H(N=24-23) and CN(N=2-1)
channel-velocity maps at $V_*=-26.5$ km\,s$^{-1}$. CN emission extends
further out than the C$_4$H emission, but, although much stronger,
is not observed at smaller radii, hence does not appear earlier than C$_4$H. 
Within the radius range where all three species are detected, the
correlation between the CN,C$_4$H and CO bright arcs is striking: this is
illustrated in the figure by the superimposition of the CN 0.025 Jy/beam brightness
contour (cyan color contour) on the C$_4$H and CO maps. As noted above, the
bright central arcs may be fitted by circular arcs whose centers of curvature
are offset by a few arcsec from the star.


Fig.~\ref{COonFORS1} shows the CO and CN (2-1) line brightness contours at the star velocity
superimposed on the VLT image of the interstellar light (V band) scattered by the
dense shell dust grains \citep{Leao2006}. Again, the 3 tracers show very similar intensity profiles 
along the arcs
(it should be noted that the imprint of a number of stars has been removed from the VLT image
so that some areas are artificially dim -- see Leao {\it ibid}). 
The agreement between CO and dust is not surprising, as the scattered light is enhanced where 
the line-of-sight is tangent to the dense shells, i.e. near the meridional plane. 
As expected, the optical arcs are somewhat broader and, at places, lie slightly inside the CO arcs.  

The similarity between the C$_4$H and CN line brightness distributions, as well as the line brightness
distributions of many other C-chain molecules and radicals (C$_2$H, C$_6$H, HC$_5$N, MgNC,...), was already
noticed in lower resolution interferometric maps by \citet{Guelin1993b} and \citet{Guelin1999}, who 
pointed out it seems to 
contradict the standard formation scheme of those species, namely carbon-chain growth initialized by
the photodissociation of acetylene. The neutral-neutral reactions involved being relatively 
slow, one may expect the species to appear sequentially at different radii. The high resolution maps 
of Fig.~\ref{COCNC4H} show that the apparition of the CN and C$_4$H emissions 
coincide in the inner envelope to better than 1$''$, i.e. a few tens of years, setting stringent 
constraints on the chemistry. This question has been addressed by \citet{Cordiner2009} and, more
recently, by \citet{Agundez2017} who stress the role of fast photodestruction reactions in shaping up
the localisation of C-chains in the envelope.

\subsection{Continuum emission}

The 1-mm continuum emission from the envelope has been previously mapped at a resolution of 11$''$ 
(HPBW) with the IRAM 30-m telescope equipped with a bolometer \citep{Groenewegen1997}. The
authors reported a broad, centrally peaked emission extending up to $r\simeq 40''$ from the star, an  
extension polluted by the telescope error beam response and by line emission. 
The bolometer band, centred 
at 240 GHz, has an equivalent bandwidth of $\simeq$ 70 GHz and covers very many molecular lines, 
some of which are very strong (CO, SiO, CS, HCN, HC$_3$N...). As to the 30-m telescope observations 
discussed in Sec. 2.3, they were carried out with HERA, a heterodyne receiver 
too unstable to yield accurate continuum information in total power mode. 

The interferometric SMA and ALMA observations have both the stability and the resolution required 
to measure weak continuum signals and discard the line emission. However, they partly 
filter out the extended emission and accurately show only structures smaller than 10$''$ 
-- the shortest spacings observed with the SMA and ALMA were 6 m and 12.7 m, respectively, 
which, despite the mosaicing observing mode, implies partial loss of smooth continuum features 
broader than 12$''$ (ALMA) and 25$''$ (SMA). 

\begin{figure}[ht!]
\includegraphics[angle=0,width=\columnwidth]{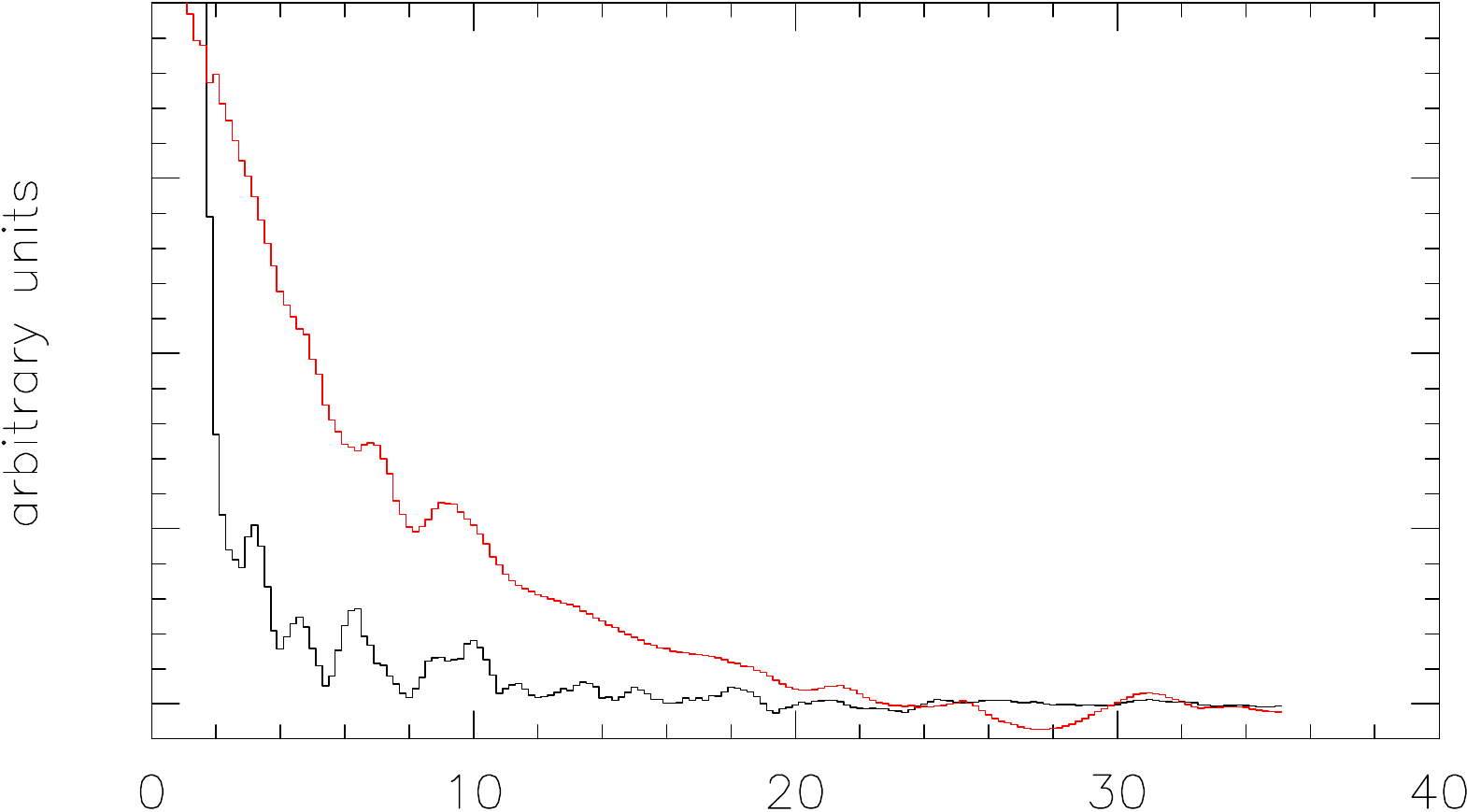}
\caption{
\label{Contavg}
Radial profile of the azimuthally averaged 1-mm continuum emission (black) and CO(2-1) emission (red).
X axis is radius in arcsec, Y axis intensity in an arbitrary scale.  The bumps correspond to segments of bright off-center arcs. 
}
\end{figure}

The continuum emission detected by both interferometers consists of a
compact central source and a clumpy halo.  A Gaussian fit of the compact
source in the ALMA uv visibilities yields a peak flux of 0.33 Jy and a FWHP of
0.22 $\pm 0.01$ arcsec. The brightness temperature of the optically
thin continuum emission is $>100$ K: obviously, the compact source corresponds to
the hot dust formation zone.

\begin{table}
\setlength\tabcolsep{3pt}  
\begin{tabular}{lrlllrl}
\hline \hline
Epoch & freq.& HPBW&RA (J2000)& DEC (J2000)& $1\sigma$ & ref. \\
      & GHz  &$''$   & $h:m:s$ & $^\circ\,:'\,''$& $mas$& $'' $ \\
2015.6 & 245 &0.3   & 09:47:57.4492 & 13:16:43.884 &4 & [1]\\
2015.92 & 650&0.3   & 09:47:57.4553 & 13:16:43.749 &20 & [2]\\
2012.29 & 260&0.6   & 09:47:57.446  & 13:16:43.86  &30 &[3]\\
2006.16 &  43&0.04  & 09:47:57.4255 & 13:16:43.815 &10 & [4]\\
1993.07 & 22.5& 0.1 & 09:47:57.39   & 13:16:43.63  &20 & [5]\\ 
\hline
\end{tabular}
\caption{\label{coord} Quoted error is 1$\sigma$ for each coordinate; references: [1]: this work; [2]: 
\citet{Decin2015}; [3]:\citet{Cernicharo2013}; [4]:\citet{Menten2012}: [5]:\citet{Menten2006}}
\end{table}

CW~Leo$\star$ is known from VLBA 7-mm
observations to have a proper motion of 35 mas yr$^{-1}$ in the NE
direction (PA=70$^\circ$) \citep{Menten2012}, which translates in
a total motion of $\sim 0.1''$ during the 3-year duration of our SMA and ALMA
observations. The position of the star in June 2015, identified as the center
of the bright 0.2 arcsec 245 GHz continuum source in our extended configuration ALMA observations, 
is 09$^h$\,47$^m$\,57.4492$^s$  13$^\circ\,16'\,43.884''$. It is adopted as the
reference position for the maps in this article. In Table 1, it is compared
to other accurate measurements made with ALMA and the VLA. We note that the 0.3 Jy continuum 
source detected at 245 GHz at the center of the map, which consists of the star plus surrounding hot dust,
is is not quite axisymmetrical. Therefore, its position slightly varies with the 
HPBW.

Fig. \ref{Contavg} shows the radial profile of the azimuthally averaged
continuum halo, compared to the azimuthally averaged CO(2-1) emission in the 
ALMA central velocity map. The clumpy continuum halo appears 
as a series of ripples of decreasing intensities, e.g. peaking near 3, 6 and 9 arcsec. 
These ripples coincide
in radius with ripples on the CO profile, which trace the bright CO arcs 
smeared by the azimuthal average. 
The ripples in Fig. \ref{Contavg} show that the innermost CO arcs are
detected in the continuum, although they are hardly visible on the
unsmoothed continuum map. After cautioning that the extended CO and continuum emissions
may be unequally filtered out by the ALMA interferometer, we note that the 
CO ripples, expressed in Jy\,km\,s$^{-1}$, are a factor of $\simeq 10$
stronger than the continuum flux density, expressed in Jy. This factor is comparable
to the ratio of the velocity-integrated CO(2-1) line flux $\int S_{CO} dv$ to the 
1-mm continuum flux $S_\nu$  observed in the cold interstellar clouds: $\simeq 8$
(see e.g. \citet{Guelin1993a}).


\begin{figure}[h!]
\includegraphics[angle=0,width=0.9\columnwidth]{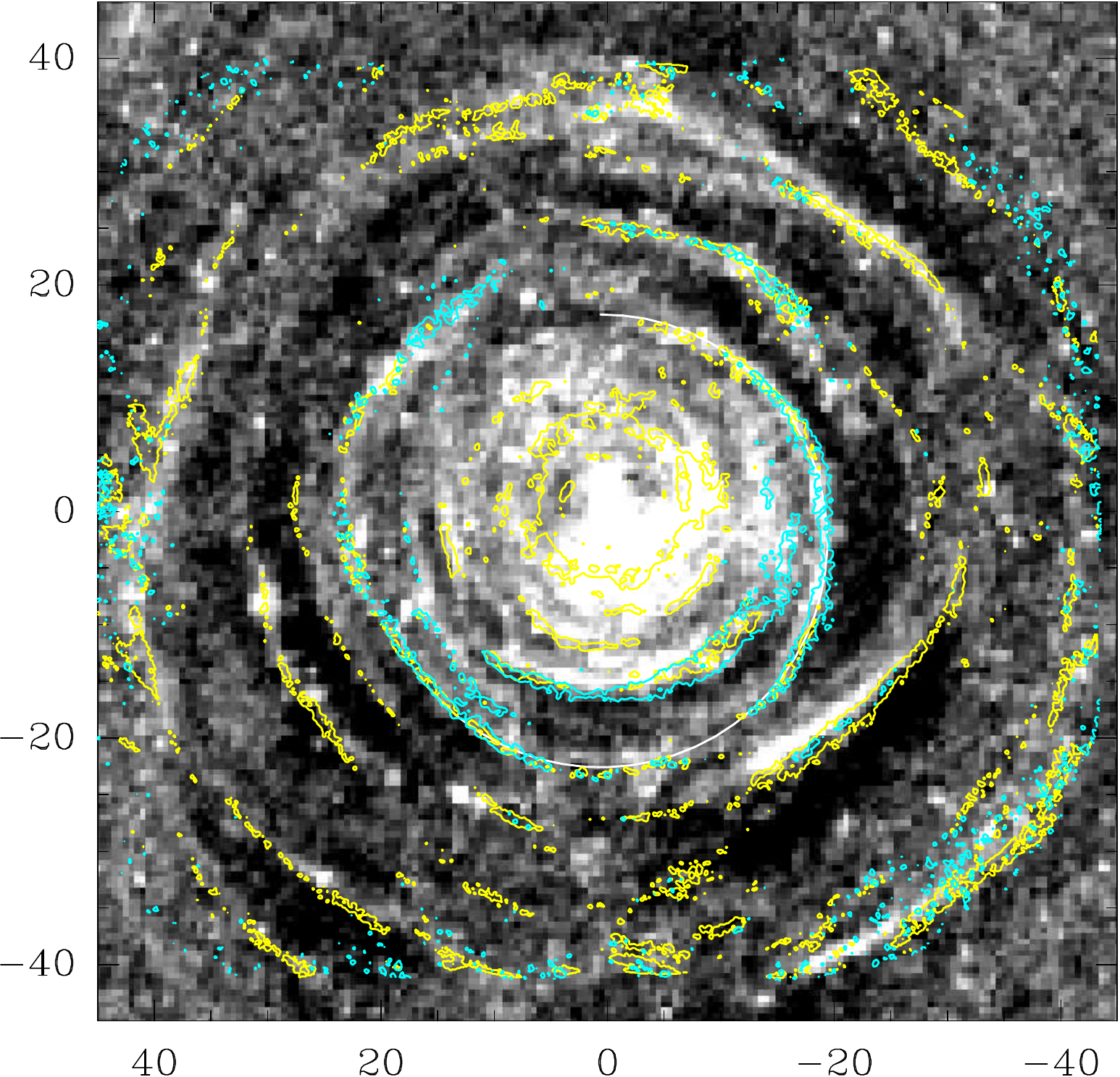}
\caption{
\label{COonFORS1}
Selected CO {\it yellow} and CN {\it cyan color} line brightness contours at  $V_*=-26.5$ km\,s$^{-1}$,
superimposed on the VLT image of scattered V light of  \citet{Leao2006}. The white circle of radius
20$''$ is centred at 0.6$''$E,2.6$''$S from CW~Leo$\star$. Coordinates are RA and DEC offsets in arcsec.
}
\end{figure}

\section{3-D reconstruction of the dense shells}

\begin{figure}[h]
\includegraphics[angle=0,width=\columnwidth]{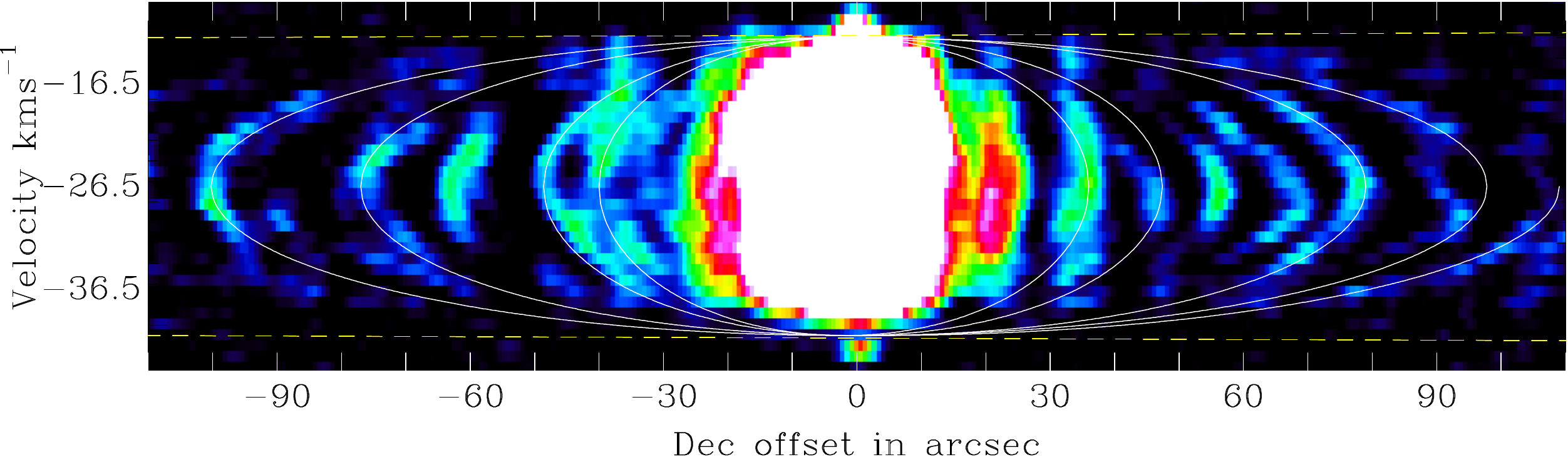}
\caption{
\label{CO-SMA-NSpvdiag} Position-velocity diagram of the CO(2-1) emission, as traced by the SMA+SD observations, 
along an 3$''$-wide $220''$-long slit oriented NS at the right ascension of the star 
($13h:16m:43.9s$). The central region of the graph
($-40'' < r < +40''$) is heavily saturated.  
The dense shells seen in Fig.~\ref{SMAv-26map} take here the form of elliptical arcs. The white ellipses 
correspond to spherical shells of radii $38'',\,48'',\,78'',\,99'', 107''$,
centred within 2$''$\, from the continuum peak and in radial expansion at a velocity of 14.5 km\,s$^{-1}$. }
\end{figure}

\begin{figure}[h]
\includegraphics[angle=0,width=\columnwidth]{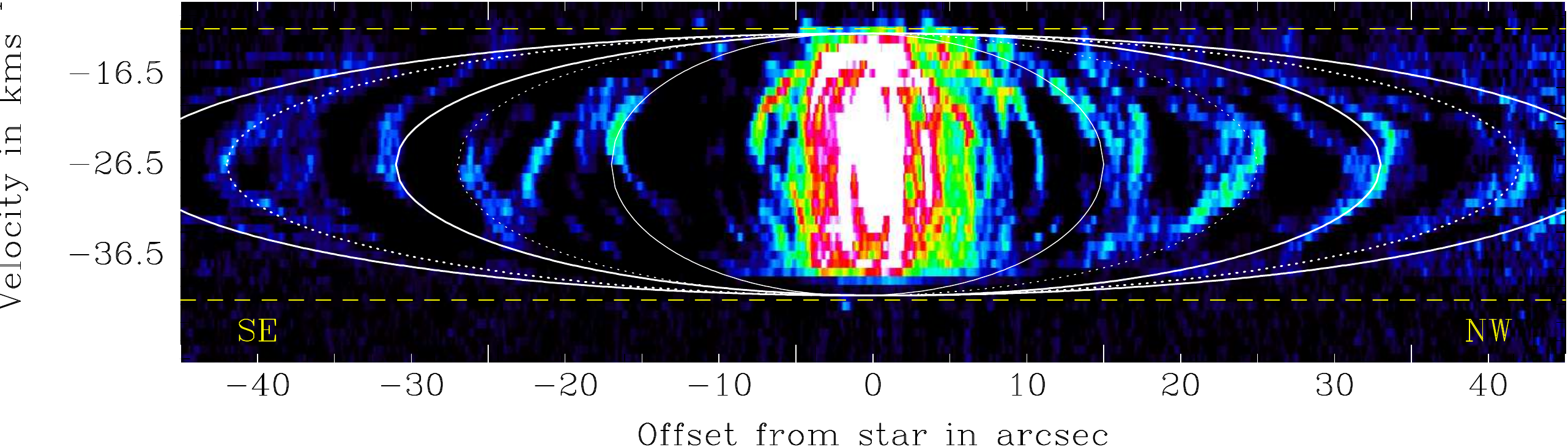}
\caption{
\label{CO-ALMA-SENWpvdiag} Position-velocity diagram of the CO(2-1) emission, as traced by ALMA observations, 
along an 0.4$''$-wide $100''$-long SE-NW oriented slit (PA=45$^\circ$) passing through 
the star. 
The dense shells seen in Fig.~\ref{SMAv-26map} appear as bright elliptical arcs. The white ellipses 
correspond to shells of radii $16'',\,32'',\,42'',\,48''$,
centred within 2$''$ from the continuum peak and in radial expansion at a velocity of 14.5 km\,s$^{-1}$.}
\end{figure}

\begin{figure}[h]
\includegraphics[angle=0,width=\columnwidth]{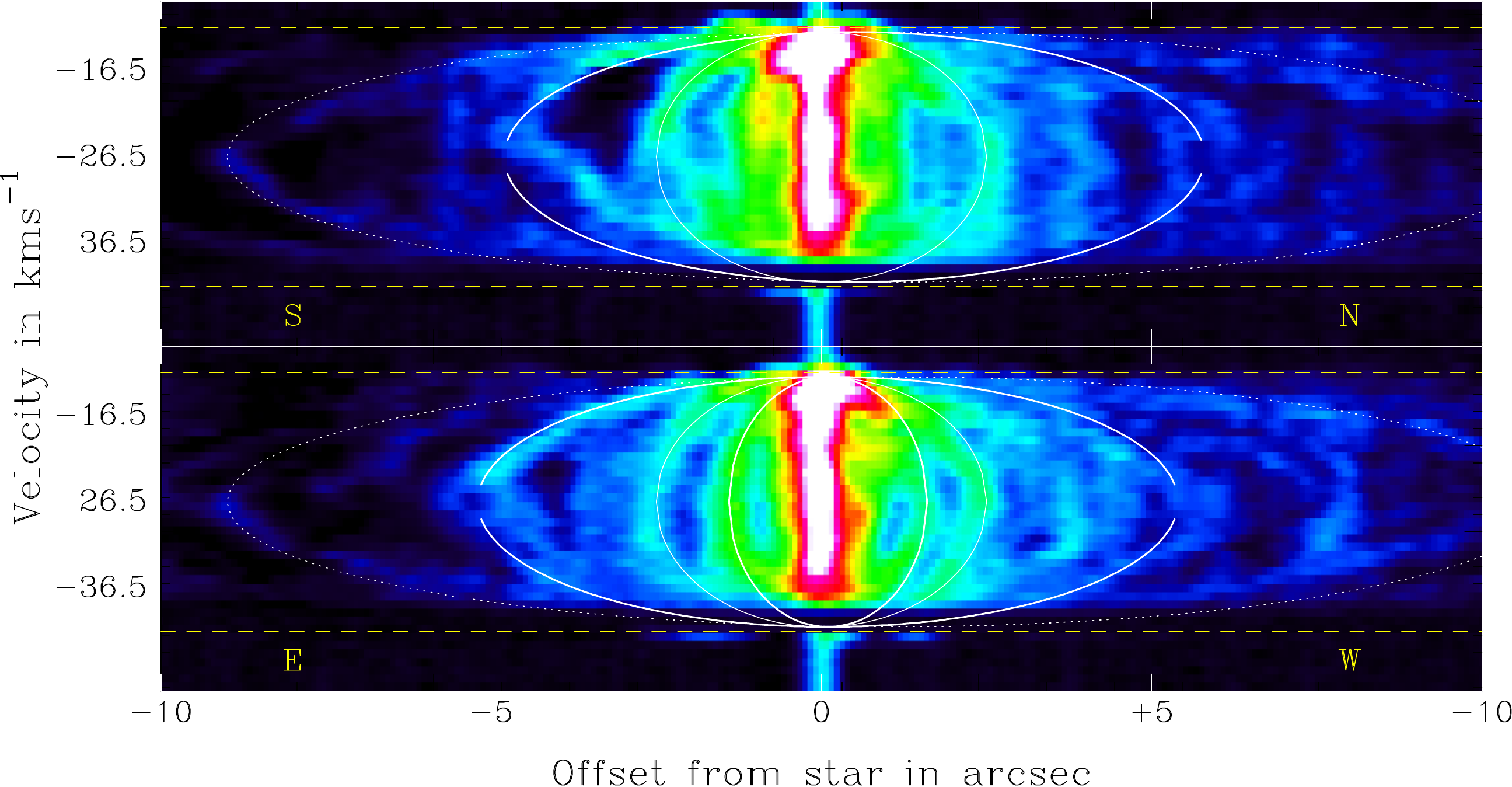}
\caption{\label{CO-ALMA-centpvdiag} Position-velocity diagram in the central region: the CO(2-1) emission, as traced by ALMA observations, 
along two 0.4$''$-wide $20''$-long slits, repectively oriented NS and EW, and passing through 
the star. 
The dense shells seen in Fig.~\ref{SMAv-26map} appear as bright elliptical arcs. The white ellipses 
correspond to shells of radii $1.5'',\,2.5'',\,5.5'',\,10''$,
centred within $0.1''-1''$ from the continuum peak and in radial expansion at a velocity of 14.5 km\,s$^{-1}$.}
\end{figure}


\begin{figure}
\includegraphics[angle=0,width=\columnwidth]{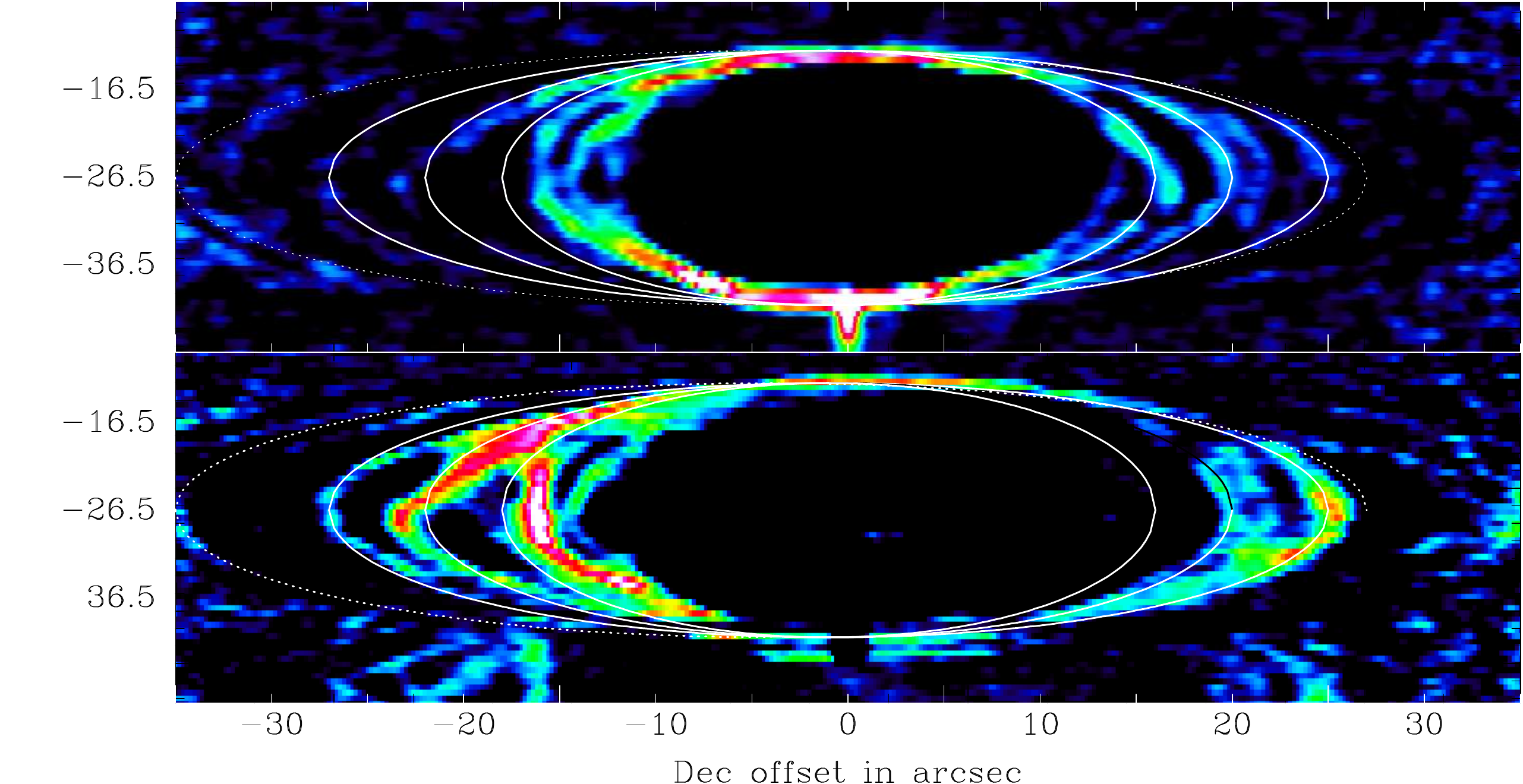}
\caption{\label{C4HNSpvdiag} Position-velocity diagrams of the C$_4$H(N=24-23) {\it (top)} and 
CN $(2_{032}-1_{021})$ {\it (bottom)} line emission along an 1$''$-wide NS slit 
at the right ascension of the star (09$^h$47$^m$57$^s$.45).  The shells on
Fig.\ref{COCNC4H} where C$_4$H is detected appear as bright elliptical
arcs. The white/black ellipse corresponds to shells of radii 17$''$,
21$''$, 26$''$ and 31$''$ centred 1$''$N from the star (4$''$N for the 31$''$ shell)
 and in radial expansion at a velocity of 14.5 km\,s$^{-1}$. Note that the CN $(2_{032}-1_{021})$ line 
shown here is blended by two weaker hfs components,  $(2_{032}-1_{022})$ and $(2_{033}-1_{023})$, respectively
11.5 MHz and 16.3 MHz higher in frequency. Those appear on the graph as ghosts of the brightest component 
shifted by -15.2 and -21.6 km\,s$^{-1}$.} 
\end{figure}

\subsection{Position-Velocity diagrams}
The discussion in section 3 was mostly based on the maps at the envelope systemic 
velocity and only dealt with the line brightness in the 
meridional plane perpendicular to the line-of-sight. We now analyze the full range of velocities
to recover the gas distribution along the line-of-sight. Let us first consider the position-velocity
diagrams along narrow slits at position angles (PA) 0, 45$^\circ$ and 90$^\circ$ and passing through 
the star CW~Leo$\star$.  

Fig.~\ref{CO-SMA-NSpvdiag} shows the position-velocity (PV) diagram of the CO(2-1) emission in the
outer envelope, as traced by 
the SMA+SD observations along a 3$''$-wide, $220''$-long slit, oriented {NS} at the right 
ascension of the star. The central part of the graph ($-40'' < r < +40''$) is
heavily saturated due the bright CO pedestal. PV diagrams of that central region, based on the 
ALMA data, are shown on Fig.~\ref{CO-ALMA-SENWpvdiag} and Fig.~\ref{CO-ALMA-centpvdiag}. In all these 
figures, 
the dashed horizontal lines correspond to $-26.5 \pm 15$ km\,s$^{-1}$, i.e. to the terminal 
velocities for an expansion velocity of 14.5 km\,s$^{-1}$, plus half the $w=$1 km\,s$^{-1}$ channel-width. 
It is striking to see how sharply the CO emission cuts-off at the {\it redshifted} terminal velocity
(the bright ridge near $\Delta \delta$ $\simeq 0$ on the SMA+SD map is residual continuum emission 
from the hot 
dust-forming region). Note that the CO emission vanishes $\simeq 2$ km\,s$^{-1}$ earlier at 
blueshifted velocities, because the cold, incoming foreground gas absorbs the bright emission from 
the warm inner layers; we show below in Fig.~\ref{C4HNSpvdiag} that the cut-off occurs symmetrically at    
$\Delta V\pm 14.5+w/2$ km\,s$^{-1}$ for the optically thin C$_4$H emission.

On the CO PV diagrams (Fig.~\ref{CO-SMA-NSpvdiag} to
\ref{CO-ALMA-centpvdiag}), the dense gas shells appear as bright
elliptical arcs, as expected for {\it spherical shells expanding at a
constant velocity}.  The white ellipses on the figures correspond to
spherical shells of radii $16'',\, 32'',\,48'',\,78'', \,99''$ and
$107''$, centred within 2$''$ from the star and radially expanding at
$V_{exp}=14.5$ km\,s$^{-1}$. Clearly, there are more bright arcs than
shown ellipses in the figures and only parts of the
model ellipses show bright emission, but overall the agreement between
observations and a series of quasi-regularly spaced shells is
remarkable. All bright arcs appear to converge toward the terminal
velocity at the star position, which implies that the expansion along
the line of sight is uniform {\it and} that the gas turbulent velocity
is small.  It should be noted that the dark vertical lane at $\Delta
\delta=28''$ on Fig.~\ref{CO-SMA-NSpvdiag} is an artifact caused by a
negative lobe in the SMA map resulting from the limited uv-plane
coverage. The lane is not present on the ALMA image, which has a higher fidelity.

The shell morphology is further illustrated in the PV diagram of
Fig.~\ref{C4HNSpvdiag}, which presents the C$_4$H and CN emission
along an NS oriented slit (very similar diagrams are observed
for the corresponding EW slits). The bright arcs closely follow the model
ellipses, $V(x)=V_{exp} \cdot cos(arcsin(x/R))$, where $x$ is the offset and $R$ the radius
of the radially expanding spherical shell. We note that the trace of
an oblate shell, resulting from addition of radial components
orthogonal to the line-of-sight, would remain closer to the terminal
velocity at low radii and fall off more briskly near the shell radius, i.e. would be more
boxcar shaped. The thinness of the arcs around the star position is
striking, allowing us to set an upper limit to
the turbulent velocity $V_{turb}$ of 0.6 km\,s$^{-1}$.

\subsection{Spherical deprojection of the outer envelope}  
We thus adopt in the following a uniform expansion velocity of
$V_{exp}=14.5$ km\,s$^{-1}$ and neglect turbulent motions. This
enables us to extract from the position-velocity CO data cubes, XYV, a
3-dimensional (3-D) spatial representation of the envelope and its
dense shells, XYZ. Two different approaches, respectively dubbed {\it Non-iterative} 
and {\it Iterative}, were used for this 3-D
reconstruction. They are described in Appendix A together with 
comparative tests on envelope models. Both yield very similar
results, though the iterative method images are less noisy and
trace the arcs closer to the star, at the expense of degraded angular 
resolution and of tight spherical boundaries. For directions close to the
center of expansion, i.e. CW~Leo$\star$, the line of sight velocity component comes to be degenerated. This,
and the resulting increase of the CO(2-1) line opacity, prevents us 
from localizing the gas cells along the 
line of sight in the approaching and receding polar caps.

\begin{figure*}[ht!]
  \centering
 \begin{minipage}{\columnwidth}
     \vspace{-0.35ex} %
     \centering
\includegraphics[angle=0,width=8.5cm]{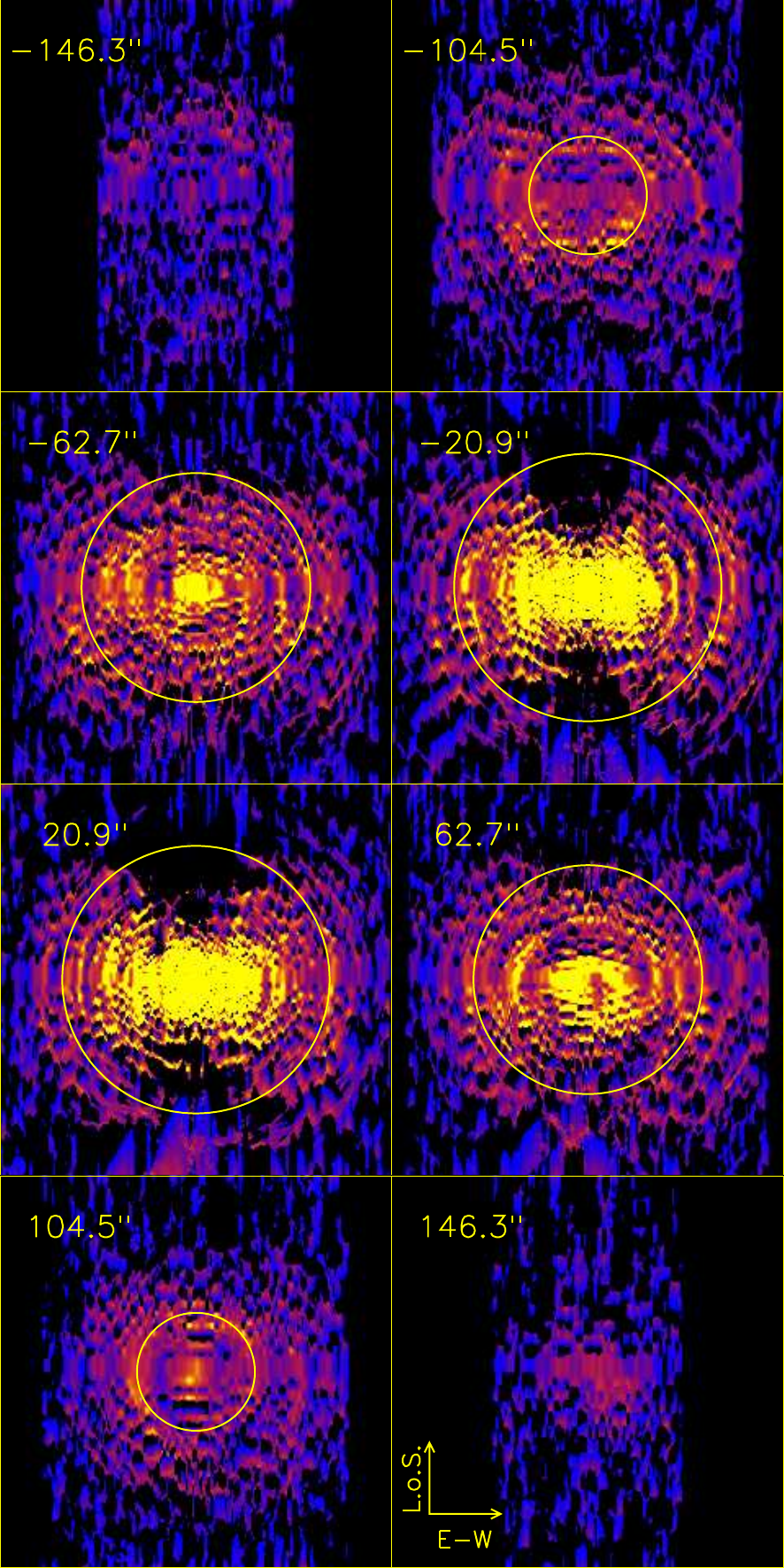}
  \caption{\label{EWplane-nonit} Reconstructed images of the CO shells {\it using the non-iterative method}. 
The images correspond to slices along 8 horizontal planes parallel to the line-of-sight (X-axis runs from 
E to W and Y-axis along l.o.s.). The slices are spaced by 42$''$; the star position corresponds to plane 
0. Image sizes are $336''\times 336''$. 
The approaching and receding halves of the envelope are down and up, respectively.
Intensity scales are identical for all images (blue= dim, yellow= bright). The areas where
reconstruction was not possible, due to velocity degeneracy or to too weak signals, are in black. 
The yellow circles mark the intersection with the planes of a sphere of radius 118$''$.}
  \end{minipage}
  \hfill
  \begin{minipage}{\columnwidth}
     \centering

\includegraphics[angle=0,width=8.5cm]{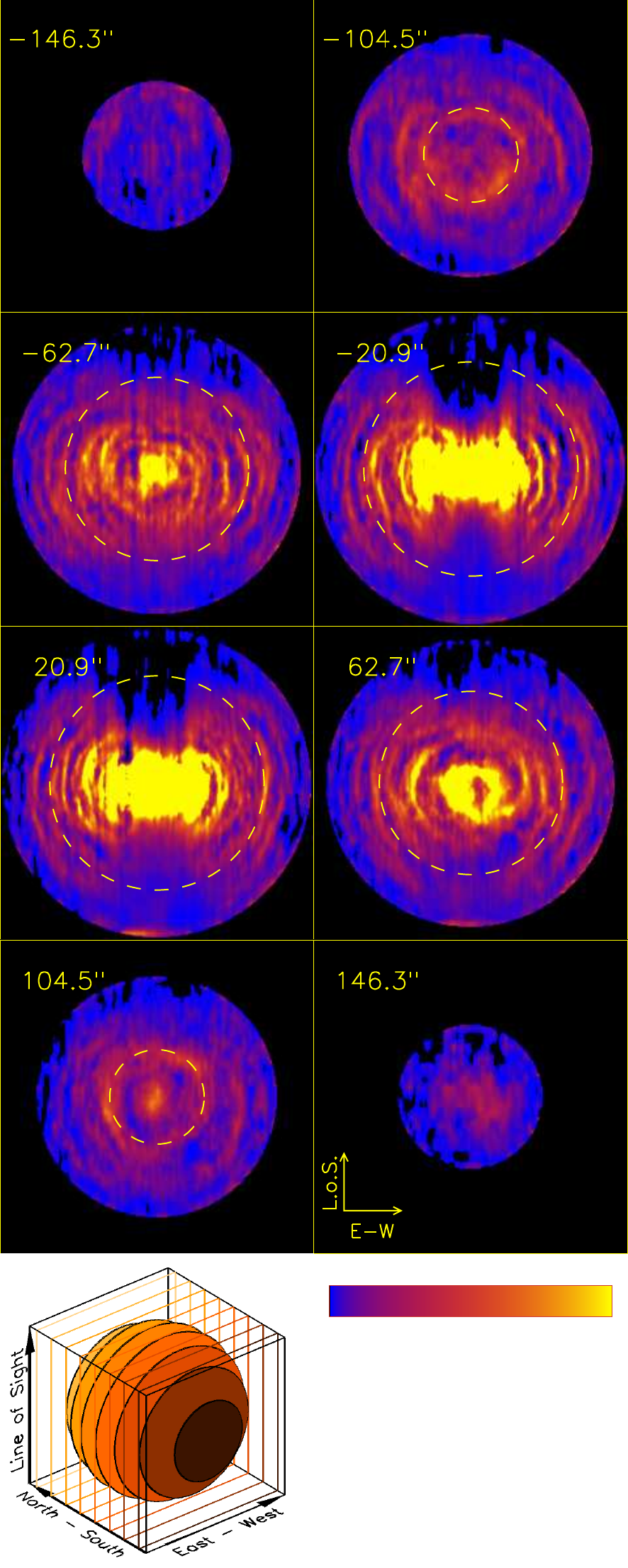}
  \caption{\label{EWplane-it} Same as Fig.~\ref{EWplane-nonit} for the {\it iterative method} (see text).
}
\end{minipage}

\end{figure*}

The results of the 3-D reconstruction of the envelope, based on the
SMA+SD data, are presented in Figs.~\ref{EWplane-nonit} for the
non-iterative method and in Fig.~\ref{EWplane-it} for the iterative
method. In both figures, the displayed frames show the reconstructed 
CO(2-1) line brightness distribution for 8 constant declination cuts 
parallel to the line-of-sight (Z axis). The planes are separated by 
42$''$ in declination (Y axis) and are located symmetrically with respect to CW~Leo$\star$. 

The bright dense shells appear in both figures as red or yellow arcs, superimposed on a
blue background. Black areas mark the regions where deprojection was
not possible because, either the line opacity is large, or the velocity  
is degenerate along the line of sight. Both conditions are met near the terminal velocities 
for the lines of sight close to CW~Leo$\star$. Moreover, at all velocities, the emission at very small radii
is dominated by the hot gas acceleration region, where our model of uniform
expansion breaks down, making the deprojection codes ineffectual.

The great similarity of the images derived by two
independent methods gives us confidence that the deprojected XYZ
representation of the envelope is unequivocal and correct.

Like the bright arcs in the plane of the sky (Fig.~\ref{SMAv-26map}),
the arcs on the maps in Fig.~\ref{EWplane-nonit} \& Fig.~\ref{EWplane-it} are close to circular
and extend symmetrically in both directions along the
line-of-sight. In frames -63$''$ and +63$''$, which are less affected by 
velocity degeneracy or by too weak a signal, the arcs form complete circles.
The yellow circles, which mark the plane intersections
with a sphere of radius $118\arcsec$, illustrate this point.

Most bright arcs can be followed when we rotate the panel planes around the 
line-of-sight by 90 degrees, i.e. until the cuts 
are oriented NS, instead of EW. We thus conclude that the 
dense shells, traced by CO in the outer envelope, have
spherical or near-spherical shapes and that the envelope is 
not manifestly oblate.

\subsection{Physical conditions and mass loss rate}

We derive the physical conditions in IRC +10 216 through a comparison of the $^{12}$CO(2-1), 
$^{12}$CO(1-0) and $^{13}$CO(1-0) or $^{13}$CO(2-1) line intensities. This was done for 
the outer envelope in our previous study \citep{Cernicharo2015}, after averaging the  
line intensities over concentric rings of width 10$''$. The lines were found to be subthermally 
excited at radii $r\geq 40''$ and the azimuthally-averaged $^{12}$CO(2-1)
line opacity $\tau$ was found in the range $2-4$; the $^{13}$CO lines are 
optically thin everywhere ($\tau <0.1$). The gas kinetic temperature was found 
to decrease from $T_{gas}= 20$ K at $r=40''$ to $\simeq 15$ K at $r\geq 60''$ and the gas 
density to decrease from  $n_{\rm H_2}= 5 \times 10^3$ cm$^{-3}$ at $r=40''$ to $1.2 \times 10^3$ cm$^{-3}$ at $r=80''$.
The latter values, coupled to a fractional abundance [$^{12}$CO]/[H$_2$] of $6\times 10^{-4}$, 
imply an average mass loss rate $\dot{M}=(2-4) \times 10^{-5} \,M_\odot$ yr$^{-1}$ from  3400 yr ago to 
1700 yr ago.

\subsubsection{Rotational and kinetic temperatures}
\label{sec:temperatures}
\begin{figure}[!hbt]
  \centering
\includegraphics[width=0.47\textwidth]{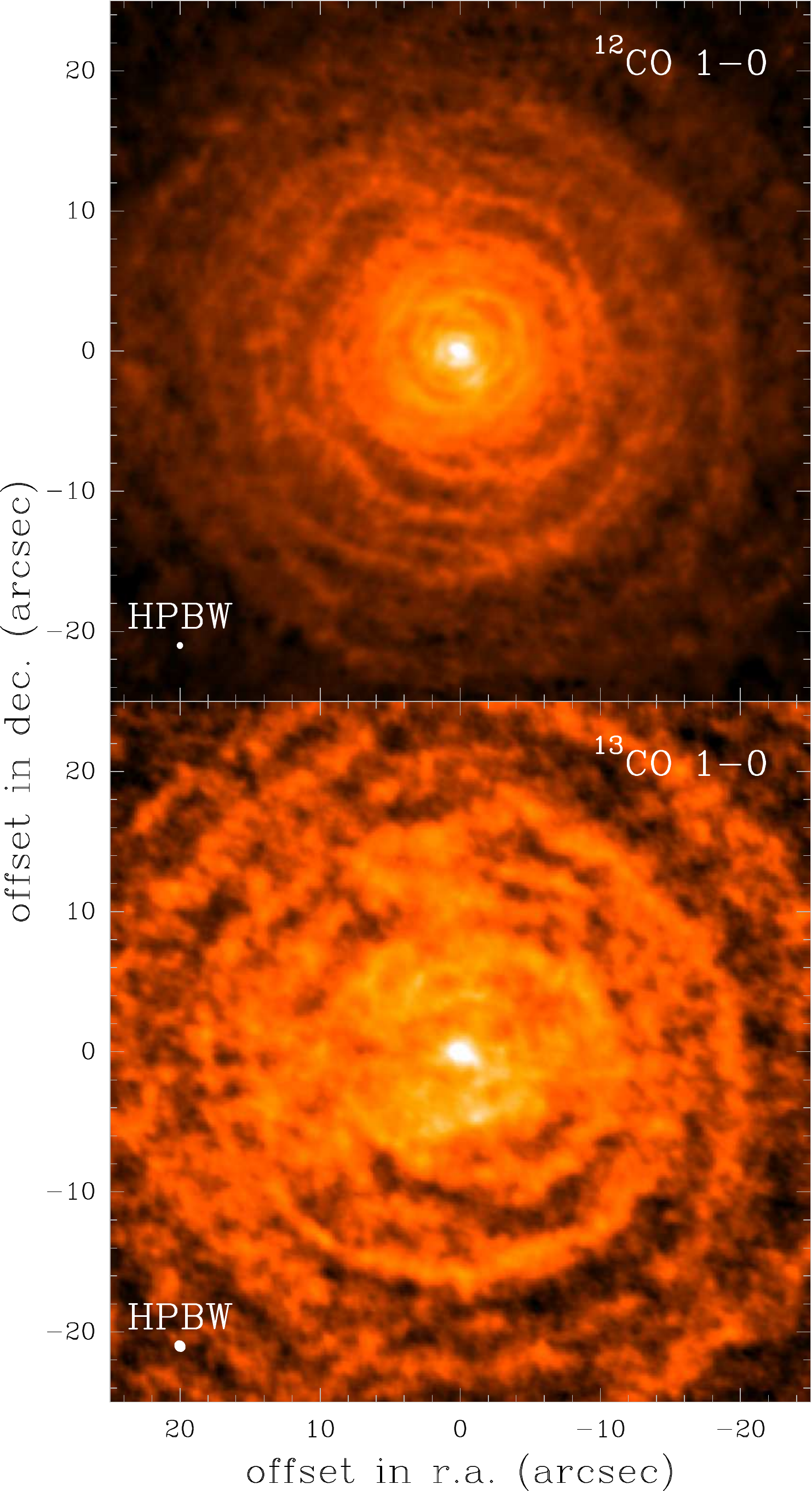}     
  \caption{The $^{12}$CO(1-0) ({\it upper frame}) and $^{13}$CO(1-0) ({\it lower frame})
    line brightness distributions at the envelope systemic velocity
    ($V_*=-26.5$ kms$^{-1}$) for the ALMA+SD data. The the synthesized beam half-power width,
    $(0\farcs48\times 0\farcs45)$ in the upper frame, has been degraded to $(0\farcs86\times 0\farcs76)$
    for the rare isotopomer line.}
\label{fig:12co10-13co10_maps}
\end{figure}

We resume this analysis for the central region, using the 1-mm and 3-mm ALMA+SD data
described in Sec. 2, more particularly the velocity-channel maps at the envelope systemic 
velocity $V_*= -26.5$ km\,s$^{-1}$. The latter provide direct information on the physical conditions
throughout the meridional XY plane. 


The $^{12}$CO(2-1) ALMA+SD velocity-channel cube (1.9 kms$^{-1}$-wide channels) was 
first smoothed to the angular 
resolution of the $^{12}$CO(1-0) cube ($0\farcs48\times 0\farcs45$ -- see upper frame of 
Fig.~\ref{fig:12co10-13co10_maps}). Then, the central ($V_*=-26.5$ kms$^{-1}$) 
velocity-channel maps of 
$^{12}$CO(2-1) and $^{12}$CO(1-0) were azimuthally averaged
over $0\farcs53$-thick concentric rings centred on CW~Leo$\star$. This allowed us to 
derive the radial profiles of the $^{12}$CO(2-1) peak line intensity, as well as of 
the $R_{21}=^{12}$CO(2-1)/$^{12}$CO(1-0) 
line brightness temperature ratio, throughout the $\simeq 50\arcsec$ central region 
Fig.~\ref{fig:12co10-13co10_maps}).

\begin{figure}[!hbt]
  \centering
\includegraphics[width=0.475\textwidth]{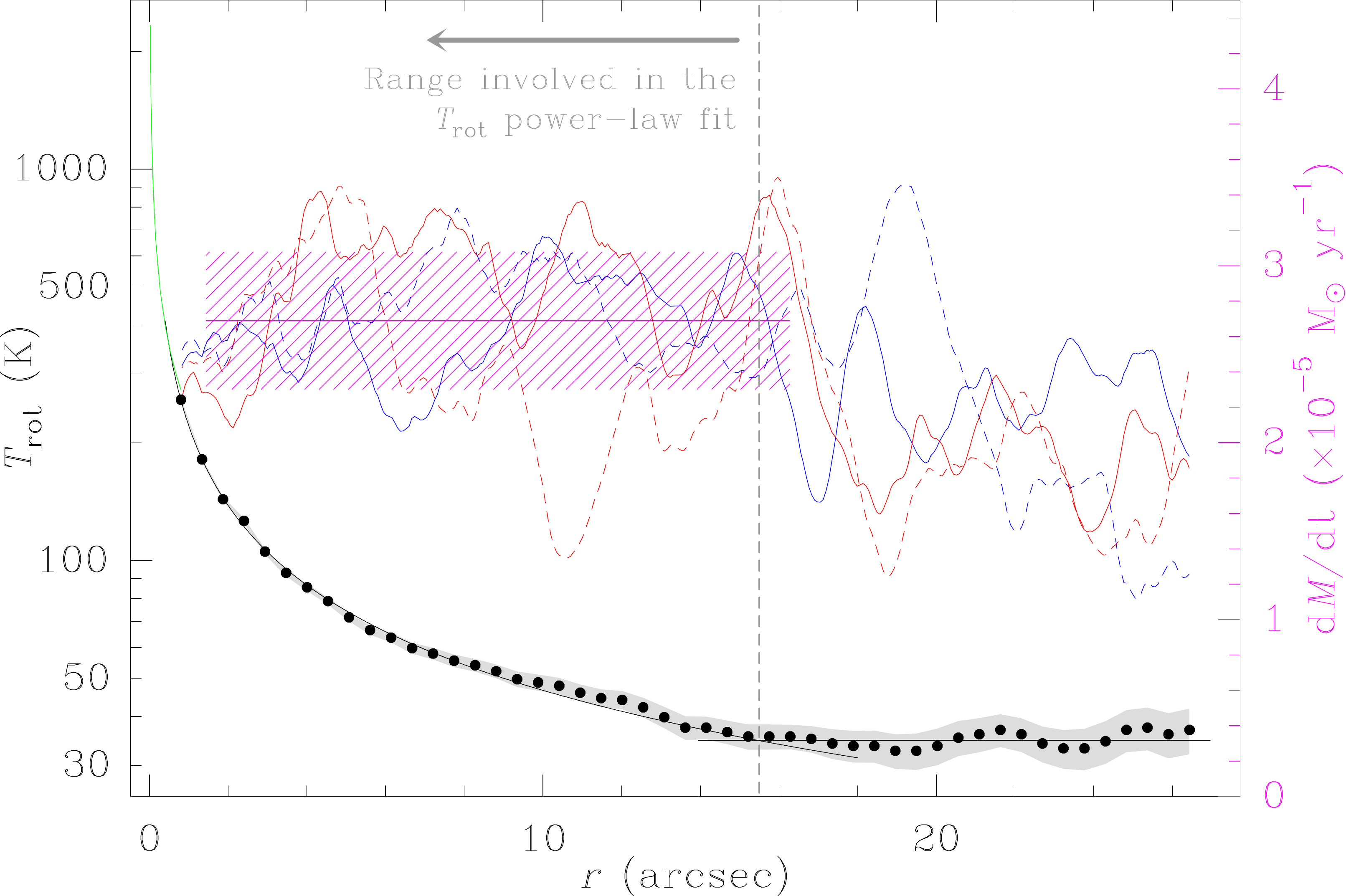}
\caption{Rotational temperature and mass-loss rate throughout the inner
  50\arcsec{} of the CSE.
  The rotational temperature (black dots) was derived from the
  $^{12}$CO($2-1$)/$^{12}$CO($1-0$) ratio averaged in concentric rings centred
  on the star.
  The green curve is the kinetic temperature close to the star
  calculated from results of previous works.
  The grey region accounts for the $1\sigma$ error interval of the rotational
  temperature.   
  The vertical dashed grey
  line indicates the distance from the star where the
  rotational temperature changes its behaviour 
  ($\simeq 15\arcsec$).
  The black continuous lines are the fits to the data set up to
  15\arcsec{} ($T_\textnormal{\tiny rot}\propto r^{-0.68}$) and beyond
  ($T_\textnormal{\tiny rot}\simeq 35$~K).
   The thin colored lines represent the mass loss rate (right axis scale) calculated 
  on the $^{13}$CO map by 
  averaging $5^\circ$-wide sectors oriented N, S, E, and W (continuous and dashed red, and continuous and
  dashed blue curves, respectively) .
The magenta
line is the average mass-loss rate and the hatched region depicts the
  typical deviation (see text).
}
\label{fig:trot_mass-loss}
\end{figure}

From the $R_{21}$ radial profiles, we derive the rotation temperature profile $T_\textnormal{\tiny rot}$ , 
which represents the kinetic temperature $T_{k}$ under the assumption that the low J $^{12}$CO levels
are at LTE. In Fig.~\ref{fig:trot_mass-loss}, the $^{12}$CO data set 
was completed in the inner $r\leq 1''$ region using the rotational temperatures derived 
by  \citet{Fonfria2015,Fonfria2017} from ro-vibrational  SiS and C$_2$H$_4$ lines.

The $T_\textnormal{\tiny rot}$ plot shows at $r \simeq 15''$ ($\simeq 750R_*$ or $2.8\times 10^{16}$~cm)
a change in dependence on the distance from the star. At lower radii, $T_\textnormal{\tiny rot}$ 
closely follows a power-law 
$T_\textnormal{\tiny rot}=(256.9\pm 1.5)\left(r/0.8\right)^{-(0.675\pm 0.003)}$~K,
where $r$ is expressed in arcsec. Beyond $r=15''$, $T_\textnormal{\tiny rot}$ remains around 35 K.

The change of the $T_\textnormal{\tiny rot}$ behaviour around $r=15''$ does not result
from a departure from the $^{12}$CO levels from LTE: modelling the $J=1-0$ and $2-1$ lines and the 
$^{13}$CO (1-0) line with the {\it MADEX} statistical equilibrium code 
\citep{Cernicharo2012},
we find that $T_\textnormal{\tiny rot}$ deviates from $T_k$ by $\lesssim 10\%$
in the range of radii considered here ($0''\leq r \leq 25''$).

Although the critical densities for the low-J CO transitions 
($\simeq 1.9\times 10^3$~cm$^{-3}$ and $1.1\times 10^4$~cm$^{-3}$, 
for J=1-0 and 2-1 lines at $\simeq 35$~K) are 
not much lower than the average gas density at $r=15\arcsec$ 
($\simeq 2.0\times 10^4$~cm$^{-3}$ -- see below),
the $^{12}$CO(2-1) line opacity  ($\tau\simeq 4$ for $r=15''$) 
prevents a significant departure from thermalization.
This is no more the case for the lines of the rare isotopomer $^{13}$CO, which, 
due to a much lower column density, are out of LTE beyond $\simeq 8\arcsec$.

Our law of the dependence of $T_k$ on distance $r$ from the star is the
first directly derived from very high angular resolution observations. 
It significantly differs from the lower resolution works of e.g. \citet{Doty1997} or \citet{DeBeck2012},
who find beyond $r=2\arcsec$ an exponent $q$ between -1 and -1.2.

Our exponent $-0.68$ is significantly shallower than that expected for 
the adiabatic expansion of a diatomic gas in the vibrational ground state,
$\simeq -1.2$. This means that the expanding gas is efficiently heated,
most likely by dust grains close to the star and by interstellar radiation
outside $15\arcsec$ \citep*[e.g.,][]{Huggins1988}.

Indeed, we know that interstellar UV radiation penetrates the envelope
as deep as $r=15\arcsec$, since the CCH and CN radicals, which form
from the photodissociation of HCCH and HCN
\citep*[e.g.,][]{Millar1994,Glassgold1996}, are first observed near
that radius (see Fig~\ref{COCNC4H}). As a matter of fact, a plethora
of radicals, such as C$_4$H, C$_3$N or MgCCH that are direct or
indirect products of photodissociation, are found to peak in abundance
at $r=14-16 \arcsec$ \citep{Guelin1993b,Agundez2017}. 

Our present observations do not have the angular resolution required
to explore the innermost envelope layers ($r\lesssim
1\arcsec$). The modelling of high energy lines arising in those layers
point to a temperature dependence to $r$ with an exponent $q\simeq -0.5$
\citet{Fonfria2008,Agundez2012,DeBeck2012,Fonfria2015}. This close to the star,
molecules are strongly excited by the radiation from the
hot dust cocoon; their collisional deexcitation is a powerful source of heat for the gas
that adds up to gas-grain collisions. Further out, the infrared
radiation gets diluted and the gas density falls off. We then
expect the exponent to steepen from $q \simeq -0.5$ to $q=-0.7$
somewhere between $r=0.5''$ and 1\arcsec.
  

\subsubsection{$^{12}$CO/$^{13}$CO abundance ratio}
\label{sec:isotopic_ratio}

In order to directly compare the $^{13}$CO and $^{12}$CO data, we
degraded the spatial resolution of the
$^{12}$CO($1-0$) central velocity-channel maps to $(0.86''\times 0.76'')$.
 
\begin{figure}[!hbt]
  \centering
 \includegraphics[width=0.475\textwidth]{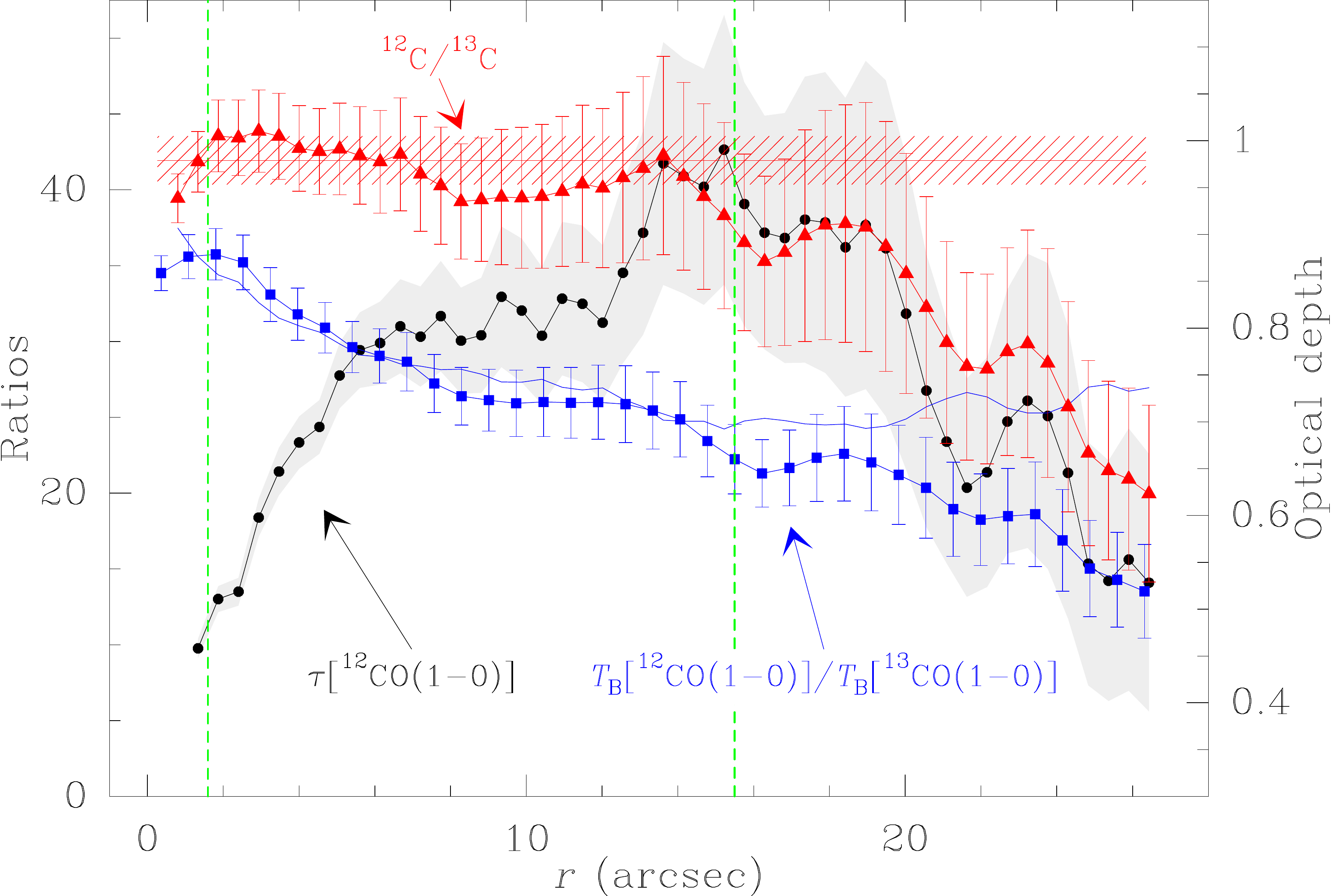}
 \caption{Ratio of the brightness temperatures of $^{12}$CO($1-0$) and
   $^{13}$CO($1-0$), $^{12}$C/$^{13}$C isotopic ratio, and optical depth of line
   $^{12}$CO($1-0$) against the distance to the star (blue squares, red
   triangles, and black circles, respectively).
  The continuous blue curve is the brightness temperature ratio
   calculated with MADEX and the kinetic temperature derived in
   Section~\ref{sec:temperatures}.
   The $1\sigma$ uncertainties have been plotted as vertical bars, a grey
   region, or a hatched rectangle.
 }
\label{fig:12c-13c_isotopic_ratio}
\end{figure}

From the resulting $^{12}$CO($1-0$) and $^{13}$CO($1-0$) line
brightnesses and the temperature profile derived in
Sec.~\ref{sec:temperatures}, we then derived  the
J=1-0 line opacity $\tau$ and the
$R_{^{12}C/^{13}C}=$[$^{12}$CO]/$[^{13}$CO] abundance ratio throughout
the meridional plane with the {\it MADEX} code. The radial profiles obtained by averaging those
quantities in azimuth are displayed on
Fig.~\ref{fig:12c-13c_isotopic_ratio}.  The isotopic ratio and the
mass-loss rate (see Section~\ref{sec:mass-loss}) have been determined
at the same time following a self-consistent iterative procedure that
can be assumed to converge after $\simeq 10$ iterations, when the
variation between consecutive iterations was
$\lesssim 0.1\%$.
We note that, contrary to $^{12}$CO, the low J $^{13}$CO levels
are found to depart significantly from LTE, with a rotation temperature
at $r\simeq 15''$ 20 to 30\% higher than $T_k$. {\it MADEX} indicates that
the difference is above 50\% beyond 20\arcsec{} from the star.


Within the errors, the ratio $R_{^{12}C/^{13}C}$ plotted in
Fig.~\ref{fig:12c-13c_isotopic_ratio} appears roughly constant from
$r=1.5''$ to $r=20''$ (the error bars on the graph only denote the
$1\sigma$ uncertainty on the line brightnesses). Its value up to the
$r=15''$ shell, $42$, agrees well with those determined from the
optically thin mm-wave lines of a number of C-bearing molecules
(C$_4$H, HC$_3$N, SiCC,...) in that very shell (40-43, see
e.g. \citet{Cernicharo2000}).

Beyond 20\arcsec, the ratio $R_{^{12}C/^{13}C}$ derived with MADEX
under the assumption that the $^{12}$CO rotation temperature equals
$T_k$, appears to decrease with distance from the star. It seems
improbable that the $^{12}$C/$^{13}$C isotopic ratio has varied in just
a few hundred years, because of a dredge-up or of chemical
fractionation. Also, selective photodissociation of CO by interstellar
UV radiation would increase, not decrease this ratio. Therefore, it is
likely that the $^{12}$CO levels start to depart from LTE for $r>
20''$, or that the rotation temperature of $^{13}$CO is underestimated in our simple model; 
for example, radiative cascades from higher-J levels pumped
by external radiation may overpopulate the J=1 $^{13}$CO level. On the
other hand, the value of $R_{^{12}C/^{13}C}$ found for radii smaller
than $20''$ and its near constancy supports our analysis and imply that
the derived line opacities and CO column densities are
accurate. Assuming the [CO]/[H$_2$] abundance ratio is known, we can
then accurately derive the mass loss rate during the last 2000 years.

\subsubsection{Mass-loss rate}
\label{sec:mass-loss}


The maps in Fig.~\ref{fig:12co10-13co10_maps} show evident arcs formed as a
consequence of the matter ejection process.
The very complex structure noticeable in these maps depicts a set of higher
density shells that are expected to be roughly spherically symmetric.
Thus, it is possible to estimate the mass-loss rate for every shell along the
line-of-sight by using the geometrical information extracted from the map of
the optically thin line $^{13}$CO($1-0$) and its rotational temperature
calculated with {\it MADEX} from the kinetic temperature derived in
Section~\ref{sec:temperatures}.

The mass-loss rate has been calculated from the brightness temperature
averaged over the position angle in four 5$^\circ$-wide sectors oriented
along the N, S, E and W directions (Fig.~\ref{fig:trot_mass-loss}).
The mean rate derived after weighting the $r\leq 15''$ data with their uncertainties is 
$\langle\dot M\rangle=(2.7\pm 0.5)\times 10^{-5}$~M$_\odot$~yr$^{-1}$.
This value was calculated assuming the $^{12}$CO abundance with respect to
H$_2$ of $6\times 10^{-4}$ estimated by \citet{Agundez2012}. The estimates beyond 
$r=20''$, which seem to show a 30\% decrease of the mass loss rate, are not reliable
in view of the uncertainties on the large departures to LTE.

The derived average mass-loss rate agrees quite well with most of the values
commonly derived or adopted in the
literature $\simeq (1.5-4.0)\times 10^{-5}$~M$_\odot$~yr$^{-1}$; see
e.g. \citet{Keady1988,Teyssier2006,Decin2010,Agundez2012,Cernicharo2015}.
The mass-loss rate radial profiles for the 5$^\circ$-wide sectors show variations by 
factors of up to 3 over scales of a few arcsec (i.e. timescales of $\simeq 10^2$ yr). This reflects the 
CO column density variation observed between the brightest arcs and the inter-arc region 
on Fig.~\ref{fig:12co10-13co10_maps}, typically a factor of 3.


\section{Discussion}

The $\sim 700$ yr time delay between the outer arcs corresponds
neither to Mira-type oscillations (the IR light period of CW~Leo$\star$ is
1.8 yr), nor to the delay between two thermal pulses ($>10^4$ yr for a
2 $M_\odot$ TP-AGB star).  In previous articles \citep{Guelin1993b,
Cernicharo2015} we tentatively explained the off-centring of the arcs
{\it and} their regular spacing by the presence of a companion star on an
elliptical orbit in the plane of the sky.  The distortion of the Roche
lobe during the companion fly-by boosts the mass loss rate at
regular intervals close to the binary period. The shells of gas
emitted at this stage drift away from the binary system in directions
that depends on the system phase at the time of their emission. Numerical simulations
--see Fig. 11 of \citet{Cernicharo2015}-- show the formation of a pattern of circular
arcs fairly similar to that observed with the IRAM 30-m telescope.

The binary star hypothesis may be supported by the detection 
with the HST of a bright red spot $0.5''$ E of CW~Leo$\star$ that
\citet{Kim2015} tentatively identify with a companion star. 
The slight curvature of CW~Leo$\star$'s trajectory between 1995 and 2001, 
recently reported by Sozetti et al (2017) from an analysis of archival
data, may be further evidence for a companion.

Analyzing $^{13}$CO(6-5) ALMA observations of a 6$''$-wide field around CW~Leo$\star$,
\citet{Decin2015} revisited this hypothesis and argued the inner 
$^{13}$CO position-velocity diagrams are better 
reproduced by a spiral structure viewed edge-on than by a
circular structure viewed face-on. In their
model, that stems from \citet{Mastrodemos1999}, the spiral
structure results from continuous mass loss in binary star system of
period 55 yr, with an orbit viewed edge-on at P.A. $\simeq 15^\circ$.  

Although the \citet{Cernicharo2015} and \citet{Decin2015} models seem
contradictory, they try to explain data relevant to different spatial scales, 
hence to different epochs. Besides, the Cycle 0 ALMA
$^{13}$CO(6-5) data, analyzed by \citet{Decin2015}, result from a
short observing time slot (17 min on-source) and are affected by relatively
poor signal-to-noise ratio and uv-plane coverage.

The data presented here have a higher sensitivity and resolution
than the previous molecular line studies and a denser uv-plane coverage,
allowing us to further investigate the origin of the CO-bright
shells.  We have seen (e.g. Figure~\ref{COCNC4H}) that the bright ring
pattern is strikingly similar for the CO, CN and C$_4$H lines at radii
where all 3 species are present. Given that the CO and C$_4$H formation paths are 
quite different and that these species abundances are poorly correlated elsewhere in 
the envelope, the regions with high line brightness must have large H$_2$ column 
densities and the arcs/filaments must trace high mass-loss events. This 
is corroborated by the excellent positional agreement between the molecular arcs 
in the meridional plane and
the dusty arcs traced by diffused IS light (Fig.~\ref{COonFORS1}). The arc/inter-arc contrast
is typically $3$ for the V-band brightness intensities \citep{Mauron2000} and $2-3$ 
for CO, which, as we have seen, implies gas and dust density contrasts close to 3.

\begin{figure}[!hbt]
\includegraphics[width=0.475\textwidth]{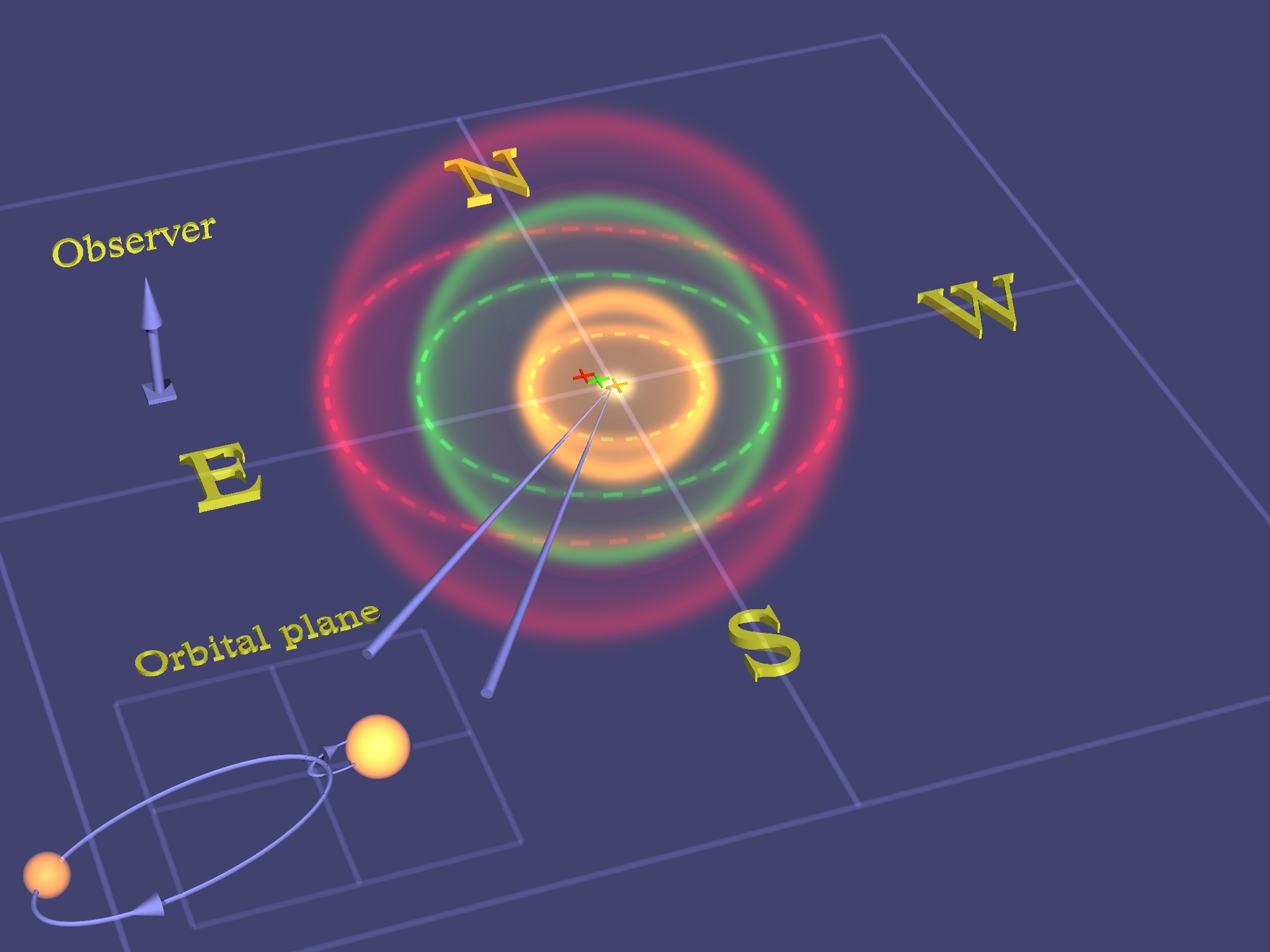}

 \caption{\label{artview} Schematic view of the 3 brightest shells of the IRC +10 216 envelope and (at an enlarged scale) 
of the binary star system with a mass ratio 4:1 and eccentricity 0.92, orbiting in the plane of the sky. The figure was generated with the POV-Raytracer 
3.6.1 software (Persistence of Vision Pty. Ltd., Williamstown, Victoria, Australia, http://www.povray.org/. 
 }
\end{figure}

We have shown in Sec.~4 that the bright arcs in the outer envelope
are not narrow structures confined to the plane of the sky, but parts of 
dense spherical shells pertaining to a spherical or near-spherical envelope.
The thinness of the shells and their fairly regular spacing  set stringent
constraints on the mass loss process and the 
binary star hypothesis.  We have repeated the numerical simulations of
\citet{Cernicharo2015} (based on the periodic ejection near periastron of spherical
shells, consisting of free-moving particles, superimposed on a lower
intensity continuous mass loss). Within the simple frame of
isotropic mass loss, constant expansion velocity and binary orbit orthogonal
to the line-of-sight, we found a set of parameters that reproduce the gross
features of the observed outer shell pattern 
Fig.~\ref{artview}).

The $\simeq 16''$ spacing (700 yr time delay) between the brightest outer arcs is found to decrease 
in the central and youngest part of the  envelope (with radius $r<40''$ and age $<1800$ yr
-- see Fig.~\ref{SMAinsert}).
It is 10$''$ (460 yr) in the NW quadrant between $r=10''$ and $r=40''$ ($5''$ 
in the inner SE quadrant) and only $\simeq 2''$ (90 yr) within 10$''$ from CW~Leo$\star$. 
Are those different patterns generated by a single binary star system and could they 
denote a rapid change of orbital period during the last 2000 yr?
Since CW~Leo$\star$ is in a phase of rapid mass loss, let us consider the effect of mass transfer 
{\it a} to the envelope and {\it b} to the companion. 

{\it a)} The isotropic ejection by CW~Leo$\star$ of an amount of matter $-dM$ into the envelope, i.e. beyond 
the binary orbit, reduces the binary system mass $M=M_1+M_2$ and tends to {\it increase}
the system orbital period (from $T$ to $T'=T\,[M/(M-dM)]^{3/2}$). However, the present rate of mass 
loss to the envelope ($\dot{M_1}=2.7 \times 10^{-5} M_\odot$ yr$^{-1}$) is 
too small to noticeably affect the period over just a few thousand years.

{\it b)} A mass transfer $dM$ from CW~Leo$\star$ to a {\it less massive} companion ($M_2<M_1$), on the other hand, would {\it decrease} 
the orbital period (to first approximation from $T$ to  $T'=T\, (M_1\,M_2/[(M_1-dM)(M_2+dM)]^3$), hence increase the frequency of fly-bys, the sought effect.

The present mass of CW~Leo$\star$ may be estimated in the range $0.6-1.2 M_{\sun}$. 
This stems from its main sequence mass $1.6 \, M_{\sun}$
derived from the envelope isotopic composition \citep{Guelin1995,DeNutte2017}, and from different evaluations of the 
envelope mass -- between $0.3\,M_{\sun}$, the gas mass expelled during the last 10$^4$ yr \citep{Cernicharo2015}, 
and $1 \,M_{\sun}$, the mass expelled since the formation of the astrosphere, 7 10$^4$ yr ago 
\citep{Sahai2010}. The  $1 \,M_{\sun}$ figures assumes the mass loss rate linearly increased by a factor of 10 during 
that time interval -- see Fig. 19 of \citet{Steffen1998}. 

Assuming the  mass of CW~Leo$\star$ is $M_1=1\,M_{\sun}$ 
and that of its companion is $M_2=0.2\,M_{\sun}$ (red dwarf of class M4V), the transfer of 0.1 $M_{\sun}$ 
from CW~Leo$\star$ to the companion would shorten the orbital period by almost a factor of 2. 
However, the transfer of $dM=0.1\,M_{\sun}$ during the last 2000 yr would imply a rate of mass transfer to the red dwarf twice 
larger than to the envelope.

The companion cannot be very massive and the mass transfer should have started 
only recently, since we see in the outer envelope no traces of direct gravitational 
interaction with the companion. The shape 
of binary star envelopes has been investigated by \citet{Kim2012, Kim2017} 
for various expansion velocities, stellar masses and orbit sizes and ellipticities. 
A near-spherical symmetry implies the outer
envelope was predominantly shaped by an isotropic wind at $V_{exp}$
and, to a lesser degree, by the orbital motion of the mass-losing star ($V_o$, 
the so-called {\it reflex} motion). A third envelope shaping
mechanism, the tidal wake generated by the companion's motion, 
should have created a flattened oblate structure, which is not observed. 
This means that $V_{exp}>>V_o$, $M_2< M_1$ and/or that the distance between the 
stars is large, which is consistent with our finding that 
$V_{exp}=14.5$ km\,s$^{-1}$, $V_{o}\simeq 2$ km\,s$^{-1}$ according to the 
observed outer arc offsets, and $T\simeq 700$ yr.

In summary, a possible scenario that explains both the outer and inner envelope morphologies, 
is then that the shells are ejected during a short period of interaction between CW~Leo$\star$
and a low-mass companion on a long period eccentric orbit.  According to
\citet{Moeckel2010}, \citet{Kouwenhoven2010} and references
therein, binary systems of large separation can form during the
dissolution phase of an open cluster, due to late captures. About 15 \%
of the binaries belong to this group, a fairly large probability. Although, it was thought
that such eccentric systems would necessarily evolve toward stable circular orbits,
the observation of eccentric companions in planetary nebulae (PNe) shows that such systems may 
remain as such for a very long time \citep{Jones2017a}.

The companion on an eccentric orbit may come very close to CW~Leo$\star$ and 
strongly interact with it. Near periastron, CW~Leo$\star$'s extended atmosphere (its
photosphere radius is 2.5 AU \citet{Fonfria2015}) may exceed the Roche limit, which
shrinks as the companion comes closer. This would raise a near-spherical shell of gas
up to the dust condensation radius (to be later accelerated by radiation pressure and 
expelled into the envelope) and, in parallel, cause a burst of mass transfer to the companion.  
In an evolved phase that may have started a few thousand years ago, the 
companion may even pass through the  upper atmosphere of the AGB star, accrete mass 
and experience drag, while resisting evaporation 
if its mass is sufficient -- e.g. $>10^{-2} \, M_{\sun}$ \citet{Soker1998}. 
The densification of the bright shells observed between the outer and the inner envelope, and the
apparent increase in the frequency of dense shell ejections,  
would then be due to a gradual reduction in
excentricity and orbital period of the companion.

Another way to explain the change in spacing between the outer and inner ejected shells,
would be to speculate that the orbit eccentricity is small and that
the star separation remains near its minimum during a fair fraction of
the period. This may give rise to multiple short-duration mass loss
episodes in the course of a single fly-by. The thin shells expelled
during such events expand quickly, while keeping the momentum
induced by CW~Leo$\star$'s orbital motion, and drift away from the
binary in different directions. The spacing and centring of the shells
in a time scale of few hundred years would then be irregular, an effect
that smears off at later times.

Alternately, we may consider that the outer shell pattern is regulated by
shocks in the expanding gas.  The expansion velocity is much higher
than the sound velocity in the envelope, hence is prone to create
shocks when the gas velocity is not quite uniform, or if the gas cannot
freely escape into outer space. This is the case, in particular, when
the gas velocity in the envelope is modulated by the orbital motion of
a binary system. The simulations of \citet{Kim2012} of a continuous
isotropic outflow of large velocity (their model M6) show that dense shells
are created where the outflowing gas, accelerated by the star orbital
motion, catches up to the gas expelled when the star was moving in
the opposite direction. The difference in velocities, 2$V_{o}$, is
larger than the sound velocity $\simeq 0.4$ km\,s$^{-1}$, so that a
strong shock occurs. The density increases (and the velocity
differential decreases) from upstream to downstream of the shock
-- the predicted increase (decrease) is a factor of $\simeq 4$ at large
radii, similar to the observed gas density contrast (see Sec. 4.3.3).

We have searched for signatures of shocks in the velocity field near the arcs.
From the CO, CN and C$_4$H PV diagrams (e.g. Fig.\ref{C4HNSpvdiag}) we see no traces
of a 2 km\,s$^{-1}$ spread in the line-of-sight velocity component, but, of course, have no
kinematical information for the perpendicular directions. If the eccentricity of the shells is caused by
a binary star system, the orbital plane should be almost perpendicular to the line-of-sight.  
 
Finally, a third possibility would be that the star system is not binary, but
triple, allowing multiple periods and different interaction configurations. 
Whereas triple stars are common in early stellar phases, 
the two closest stars tend with time to form a common-envelope and to merge. Yet, 
\citet{Soker2016} argue that triple star systems may survive up to the AGB stage: 
from the images of hundreds of PNe they find that one in six seems to host such a 
system. However, as pointed up by \citet{Jones2017b}, so far only one PN has been 
actually demonstrated to host 3 stars at its center, on the basis of velocity variations. 


\section{Summary and Conclusion}
 The main findings of our IRC +10 216 observations can be summarized as follows:

1) The star CW~Leo$\star$ and its hot dust cocoon appear in the 1.2 mm continuum 
as a 0.33 Jy Gaussian source of diameter 0.21$''$. 

2) In the CO(2-1) line, the envelope consists of a strong, fairly compact source, 
centred on the star, plus a slowly decreasing component, extending up to $r=3'$ 
from CW~Leo$\star$ and a series of bright shells that modulate the extended component. 

3) Outside the dust acceleration region, the envelope expands radially  
at a remarkably 
constant velocity, 14.5 km\,s$^{-1}$, with a small turbulent velocity ($\leq 0.6$ km\,s$^{-1}$).

4) At the star velocity,  i.e. in the plane of the sky passing through  
the star, the bright shells trace a pattern of circular arcs as thin as the  
3$''$ SMA beam. Outside radius $r=40''$, the brightest arcs are nearly concentric and  
regularly spaced, with an average separation of 16$''$, or 700 yr in expansion time. They are  
centred a few arcsec on either side of CW~Leo$\star$.

5) Inside radius $r=40''$, ALMA reveals a denser, less regular arc  
pattern. The typical separation between the brightest arcs decreases, a change that can  
hardly be attributed to the tenfold higher resolution, since smoothed to 3$''$, the ALMA arcs  
follow the SMA inner arcs. ALMA resolves the bright arcs into thin circular filaments  
off-centred from CW~Leo$\star$ by a fraction of an arcsec for the smallest arcs, up to  
a couple of arcsec for the largest ($20-30''$). Some filaments west of the star, appear kinked.

6) The pattern of bright CO arcs is accurately replicated in the CN  
and C$_4$H line emissions, wherever theses radicals are present, as well as in the dust  
structures traced by scattered light. They must trace gas shells $\simeq 3$ times denser than the intershell  
regions.

7) The gas kinetic temperature in the central region follows 
from $r=0.5''$ to $r=15''$ the power law 
$T_\textnormal{\tiny rot}=(256.9\pm 1.5)\left(r/0.8\right)^{-(0.675\pm 0.003)}$~K, and settles around 35 K 
until at least $r=25''$. The mass loss rate during the 
last 10$^3$ yr was $\langle\dot M\rangle=(2.7\pm 0.5)\times 10^{-5}$~M$_\odot$~yr$^{-1}$ on the 
average, with peak-to-peak variations of a factor of 3. The $^{12}$CO/$^{13}$CO abundance ratio
is found equal to 42 in the central region.

8) A 3-D reconstruction of the outer envelope shows the dense shells have nearly spherical shapes, 
extending typically over one or more steradians. The envelope shows little or no signs of flatness.

 The outer envelope structure (the spherical shape of the shells and  
their small eccentring) may be explained by continuous mass loss plus short mass-loss episodes  
triggered by the fly-by of a companion star with a 700 yr period and orbit nearly perpendicular to the
line-of-sight. The orbital velocity of the AGB star is  
typically 1-2 km\,s$^{-1}$ during the episodes of high mass loss. 

The structure observed in the central 
region of the envelope, where the shells are denser, less regular and show no obvious signature of a 
spiral, is more difficult to explain. It may have been shaped by the companion star described above, 
if its mass is much lower (e.g. $1/5^{th}$) than the mass the AGB star, and if a large transfer of mass 
to the companion has developed during the last 2000 yr. Alternately, the central system may consist of more than 
two massive bodies. 

Long term very high angular resolution monitoring of the compact central source
in the visible, IR or the sub-mm continuum may allow us to further investigate its nature.   
The dim source spotted by \citet{Kim2015} on HST images, and the slight curvature of  
the CW~Leo$\star$ track on the sky could be first evidence for a multiple system.

\begin{acknowledgements}

We thank S. Guilloteau and S. Bardeau for their help on the GILDAS software and
J.M. Winters for helpful comments. 
This paper makes use of data from the following instruments: The
Submillimeter Array (SMA project number 2011B-S052, 2012B-S004, and
2012B-S101), IRAM single dish telescope and
interferometer (projects t020, v018), ALMA (ADS/JAO.ALMA\#2013.1.01215.S). 
The work was partly supported by the CNRS program ``Physique et Chimie
du Milieu Interstellaire'' (PCMI). MG also acknowledges support for travel by the SMA.
We thank Spanish MICINN for funding support
through grants AYA2006-14876, AYA2009-07304 and the CONSOLIDER
program "ASTROMOL" CSD2009-00038, and the European Research Counil for funding support through
the European Union's Seventh Framework Program (FP/2007-2013)/ ERC Grant Agreement n. 610256 
NANOCOSMOS.
\end{acknowledgements}
\bibliography{references}

\section{Appendix A: Spherical deprojection of the X,Y,V CO data cubes}
A number of techniques have been elaborated to take account of the
projection effects that are a daily problem in astronomy. Other than
model fitting, deprojection recovers a spatial emissivity structure
based on {\em ad hoc} symmetry assumptions. Compared to a typical
model fit, it allows the recovery of more detailed structures
at the cost of purely geometrical rather than physical constraints;
it is well adapted to objects that by nature exhibit a prominent 
symmetry. 

Spherical deprojections based on
the Abel transform were used by \citet{vonZeipel1908} on
globular clusters.  \citet{Kaastra1989} extended the method to
cylindrical symmetry for a reconstruction of Tycho's supernova
remnant. \citet{Hubert2016} combined the Abel transform with
least-squares fits to triangular profiles to reconstruct cometary
atmospheres. However, going further than spherical symmetry makes
the results ambiguous.

The principal difference between deprojection and a tomographic 3-D
reconstruction resides in the uniqueness of the observer's viewing
angle for deprojection, uniqueness that may lead to fallacious representations.  
 \citet{Rybicki1987} argued that the Fourier slice theorem
made it impossible to recover information from inside a ``cone of
ignorance'' of half-opening angle $\theta$ for an object whose symmetry axis
is tilted by $\theta$ relative to the plane of the sky;
\citet{Palmer1994} constructed a Legendre-polynomial based expansion
for density distributions that met the condition that each single term
must not be based exclusively on terms within the cone of ignorance,
requiring an absence of abrupt changes in the density distribution. An
iterative deprojection algorithm was presented by \citet{Bremer1995},
but although the method was stable enough to support noisy images and
deconvolution it found its limits at the cone of ignorance.
\citet{Gerhard1996} demonstrate how axisymmetric density distributions
can be constructed that are invisible when observed under a range of
inclination angles; such distributions may be added and subtracted
arbitrarily from a deprojected cylindrical density distribution
without changing the projection. The infinite variety of those
functions showed that the absence of abrupt density changes was not
enough to call a distribution sufficiently ``well behaved'' to justify
its deprojection.  Only when the symmetry axis is in the plane of the
sky is the cylindrical symmetry deprojection unique.

For our deprojection analysis of the IRC +10 216, we place the symmetry
constraints on the velocity field, assuming a constant spherical
expansion velocity. As seen in Sec. 4.1, this assumption is justified 
by the shape of the bright arcs in the position-velocity diagrams and the 
similarity of the line widths for the various tracer molecules of the envelope, both 
of which point to a constant expansion velocity of 14.5 km\,s$^{-1}$.

\begin{figure}
 \includegraphics[angle=0,width=0.85\columnwidth]{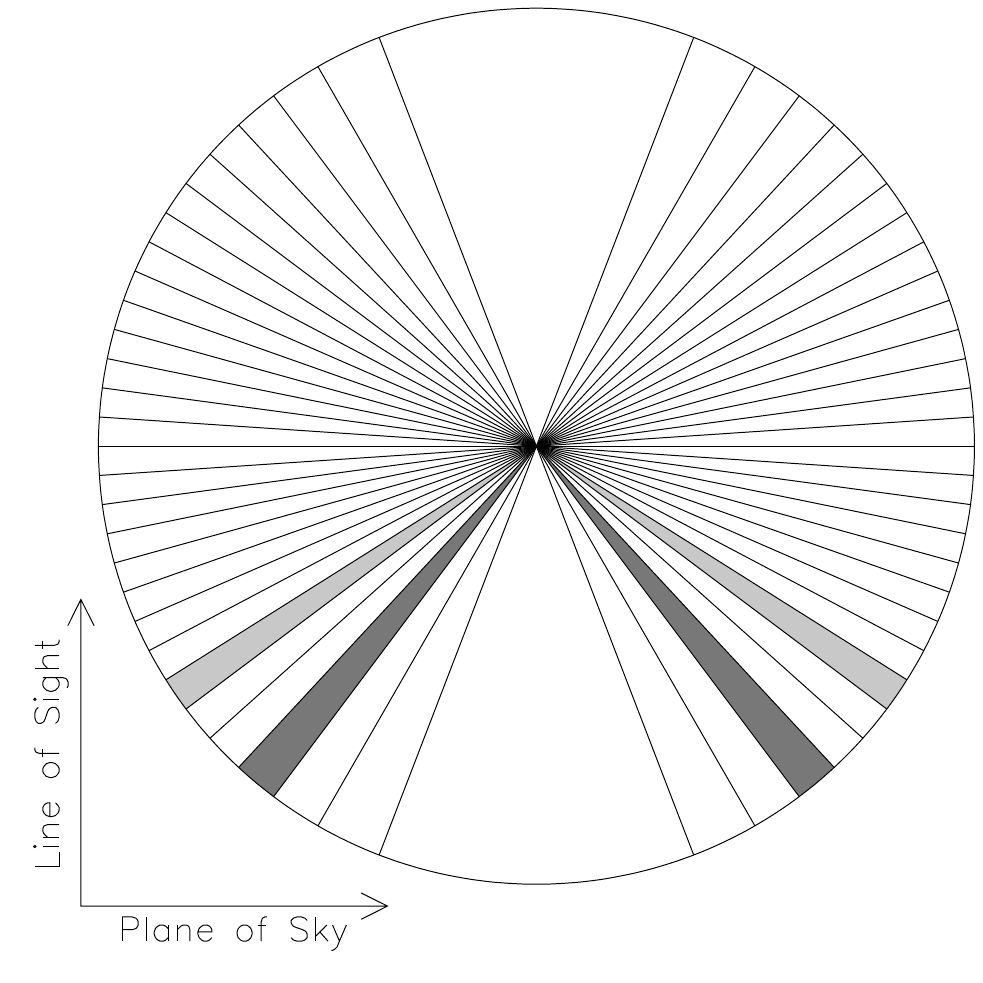}  
  \caption{\label{fig-reco-1} Volumes sampled in the velocity 
  image of a sphere expanding at constant velocity. A total of $nc=30$ velocity channels was assumed here; 
  light and dark grey represent volumes projected into the fourth and seventh channel, respectively.
}
\end{figure}

\begin{figure}
  \includegraphics[angle=0,width=0.97\columnwidth]{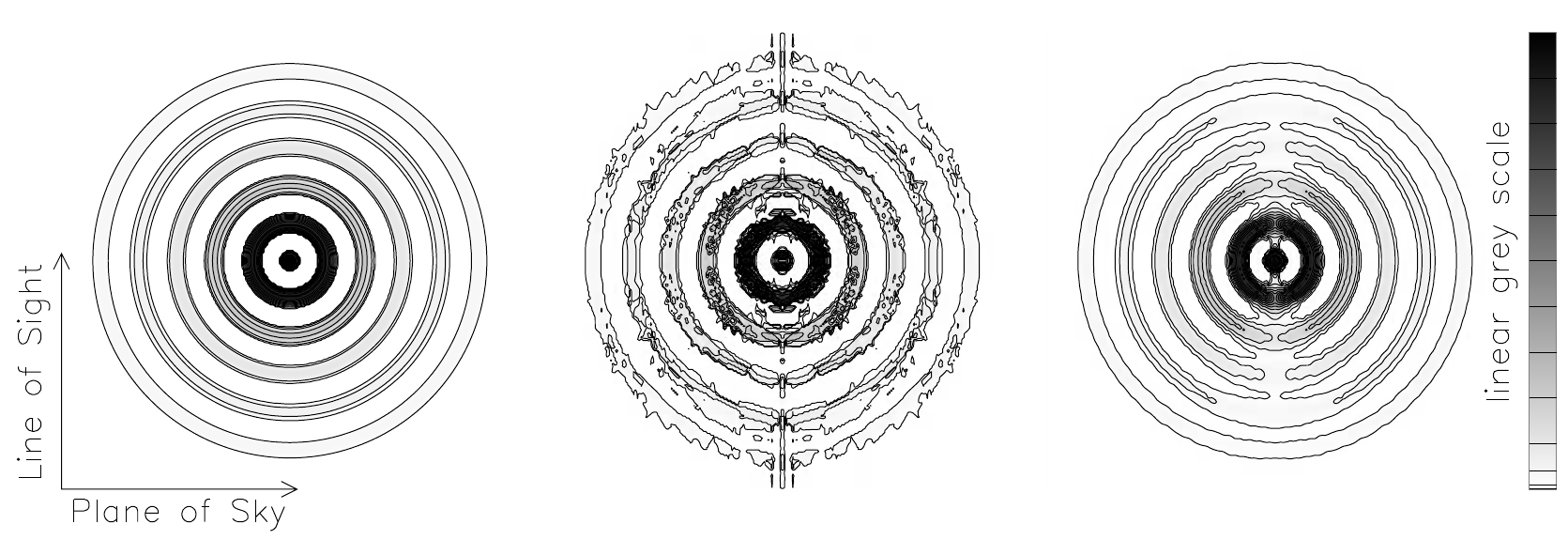}
  \caption{\label{fig-vtest-1} Case (1). Concentric shells of
    diminishing emissivity, from left to right: Synthetic
    distribution, reconstruction with method {\bf A} and finally
    {\bf B}. Intensity scales are identical for all three.  
}
\end{figure}
\begin{figure}
  \includegraphics[angle=0,width=0.97\columnwidth]{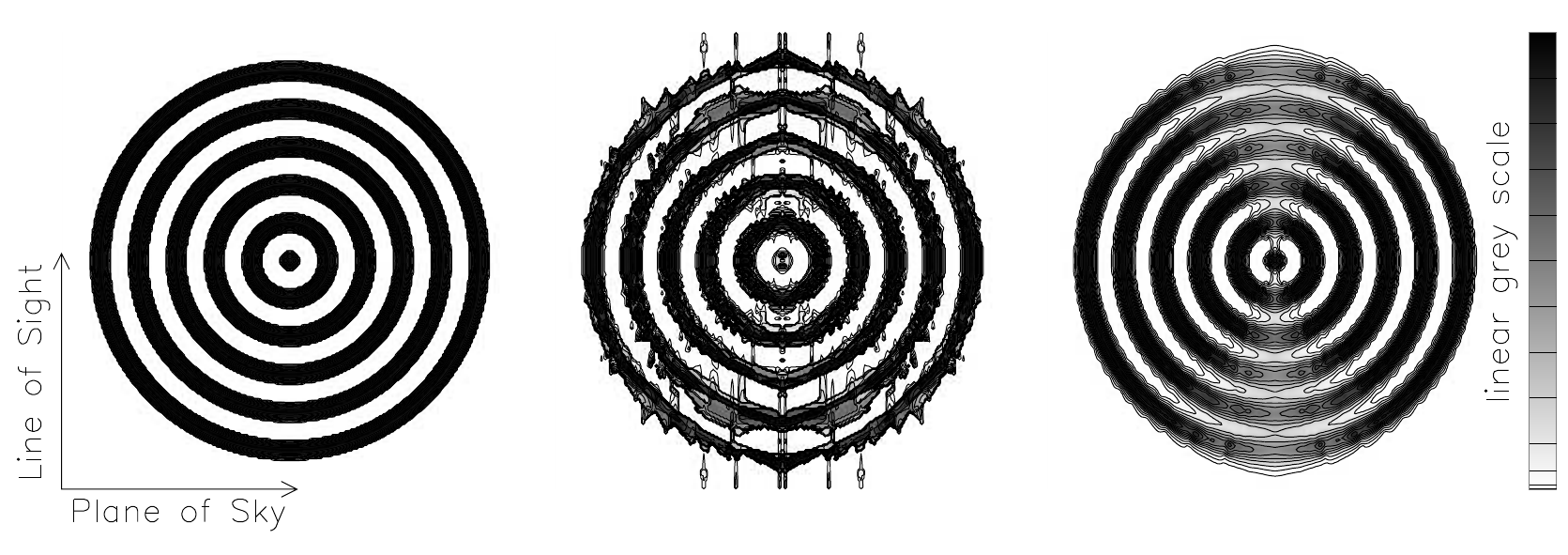}
  \caption{\label{fig-vtest-2} Case (2). Concentric shells of
    constant emissivity, from left to right: Synthetic
    distribution, reconstruction with method {\bf A} and finally
    {\bf B}. Intensity scales are identical for all three.  
}
\end{figure}
\begin{figure}
  \includegraphics[angle=0,width=0.97\columnwidth]{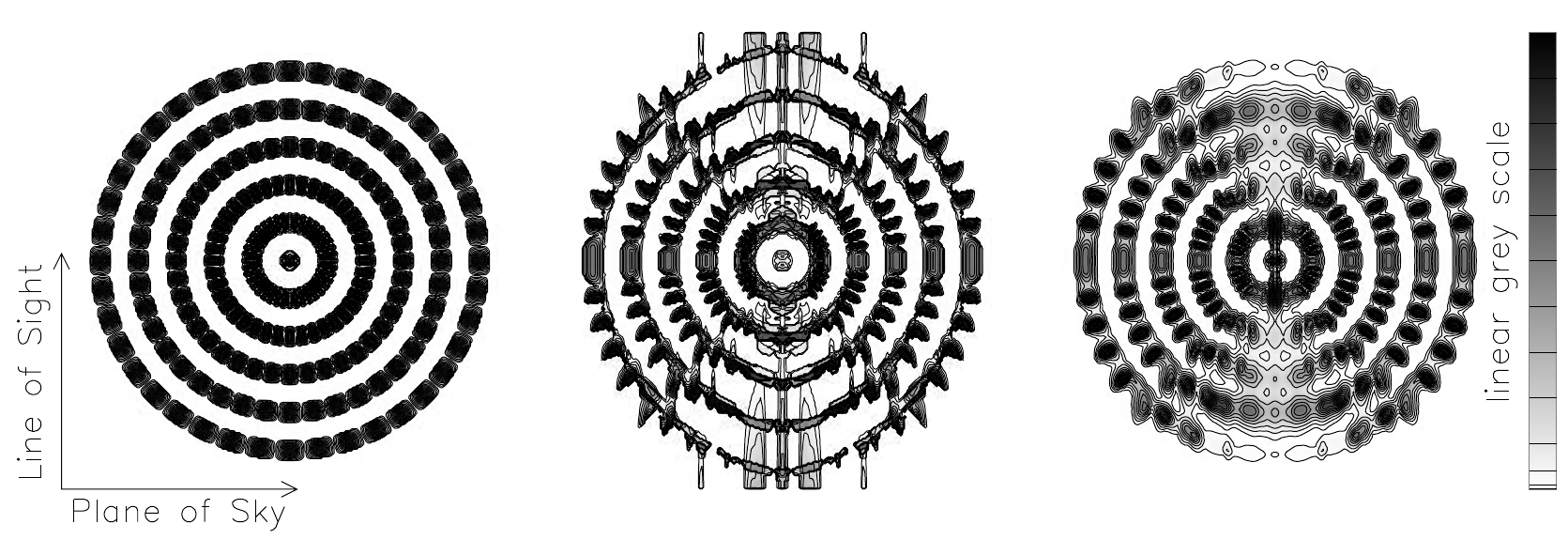}
  \caption{\label{fig-vtest-3} Case (3). Concentric shells of
    constant emissivity with a strong modulation in the spherical $\theta$ direction, from left to right: Synthetic
    distribution, reconstruction with method {\bf A} and finally
    {\bf B}. Intensity scales are identical for all three.  
}
\end{figure}
\begin{figure}
  \includegraphics[angle=0,width=0.97\columnwidth]{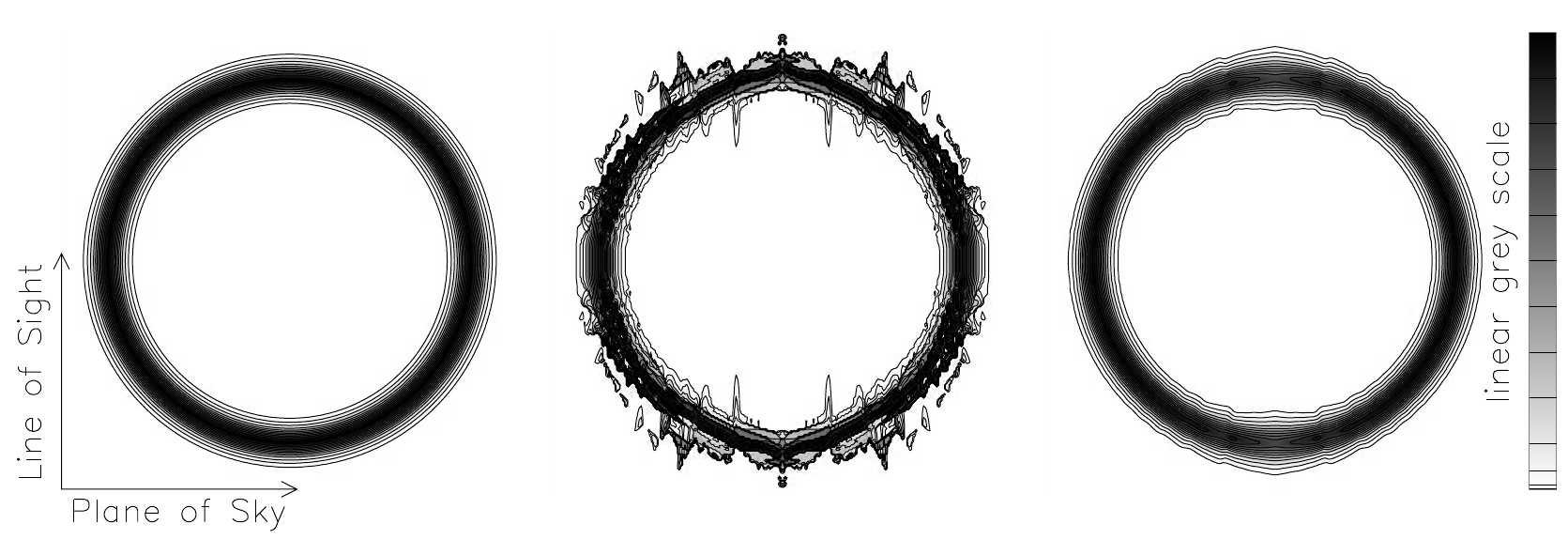}
  \caption{\label{fig-vtest-4} Case (4). A simple expanding shell, from left to right: Synthetic
    distribution, reconstruction with method {\bf A} and finally
    {\bf B}. Intensity scales are identical for all three.  
}
\end{figure}
From the position-velocity datacube (spatially resolved in x and y and binned 
in the line of sight velocity component, $V$, with steps of $\Delta V=1$ km\,s$^{-1}$) 
we first generate a set of velocity-channel maps, each corresponding
to the integral along the line of sight of the signals emitted within its velocity bin.
We note that for a spherical volume and a uniform expansion velocity, $V_{exp}$, the 
1 km\,s$^{-1}$-wide bins define spherical sectors of equal volume (see
Fig.~\ref{fig-reco-1}). Close to the line of sight towards the
observer, we have to expect a final cone of half-opening angle $alpha
= arccos(1 - 2/nc) \simeq 21^\circ$, where $nc=2V_{exp}/\Delta V$ is the number of spectral channels
required to sample the sphere. 

Starting from the (x,y,V) datacube, we follow two different approaches for the 
reconstruction of the spatial 3-D envelope morphology. Both
assume a constant expansion velocity, discarding additional motions 
in the plane of the sky that are unobservable. Both also assume that the observed CO signals
are proportional to the gas column density.
\begin{itemize}
\item {\it {\bf A}: Non-iterative code}. The code works with Cartesian coordinates. It
  starts from the
  central velocity-channel map ($V=V*$ bin) and moves
  alternately towards the adjacent external velocity bins. For every
  pixel x,y in the sky plane, it identifies from the pre-defined velocity 
  field the coordinate range along the line of sight (z direction) that corresponds 
  to each velocity bin. For the first bin ($V=V_{*} \pm \Delta V/2$), it evenly distributes 
  the emission observed toward each x,y pixel over all cells in the allowed z segment. 
  Turning to the next velocity bin, it first sets the emissivity in 
  each x,y,z cell of that velocity range by taking the average value of its closest known 
  neighbours. Then, it calculates for every x,y the integral of this first guess over z 
  and compares it to the observed signal in the velocity-channel map. The difference 
  guessed-observed, weighted by the square of the distance to the closest neighbours, is 
  subtracted from the guessed values, starting from the most distant (highest weight) cell to the
  closest cell; if the subtraction yields a negative value, the latter is set to zero 
  and the value to be subtracted carried over to the next z cell. The main time consuming 
  feature of this code is the distance weighting and the sorting.

\item {\it {\bf B}: Iterative code}. This code switches back and forth from Cartesian to spherical
  coordinates. It first {\it uniformly} spreads the emissivity observed in every velocity-channel 
  map for each x,y pixel over all z cells allowed for that bin. Next, it resamples by linear 
  interpolation the so derived emission model onto a spherical coordinate grid ($r$, $\phi$, $\theta$). 
  Then it performs in spherical coordinates a series of smoothing steps along the polar 
  ($\theta$) angle with a weighting that tends to diffuse the signal toward the polar regions. Finally,
  it interpolates back the emission into the Cartesian 3-D grid, calculates the resulting 
  velocity-channel image and normalizes the result to match the input image. The interpolation/diffusion
  steps are repeated until a stable solution is obtained.
\end{itemize}

The density reconstruction algorithms run into three problems, all the more severe 
when close to the incoming/receding polar caps:

\begin{itemize} 
\item {\it Line opacity}: The $^{12}$CO(2-1) line opacity in IRC +10 216 is fairly large. It  
   is typically 1-2 in the central velocity-channel and reaches values up to 3-4 between $r=10''$ and $20''$ 
  (see Sec. 4.3.2). In the outer($r>40''$) envelope, however, the line become subthermally 
  excited, so its intensity roughly remains proportional to the CO column density, validating our 
  above approach. Such is not the case, however, near the extreme velocities, where the line opacity
  becomes much larger. 
 
\item {\it Degeneracy}: Velocity degeneracy does not allow to fully reconstruct the 
  emission along the lines of sight crossing the cone-shaped regions with the 
   most extreme velocities. The algorithms only do an extrapolation, which is tested for
  consistency against the velocity image.

\item {\it Gridding}: The innermost part of the envelope is the brightest part in
  $^{12}$CO(2-1). Its small-scale structure would require a denser
  gridding in the calculation of the velocity image.
\end{itemize}

This explains why the 3-D reconstructions of the CO envelope have difficulties
closing the dense shells near the poles. The position-velocity diagrams of other 
molecular gas tracers, e.g. the optically thin C$_4$H lines, strongly suggest that 
the shells do close near the line of sight
to the observer. We can trace the shells over large angular ranges of
about 90 degrees outside the plane of the sky (see e.g. Fig.~\ref{EWplane-it}) 
and about 120 degrees inside (Fig.~\ref{SMAinsert}). The morphology is rich in narrow spurs 
bridging the shells, which may indicate
remnants of shells emitted at slightly different phases in the orbit. The degree of interaction
between the shells is not strong enough to produce a telltale
broadening of the observed spectral lines.

Both methods have their advantages and drawbacks, which we demonstrate
with four numerical testcases in Fig.~\ref{fig-vtest-1},
Fig.~\ref{fig-vtest-2}, Fig.~\ref{fig-vtest-3} and
Fig.~\ref{fig-vtest-4} that were calculated under conditions matching
the observations for IRC +10 216. Code {\bf A} allows a better
conservation of the contrast between different shells at the price of
a higher noise level in the reconstructed cube, while code {\bf B}
produces smoother distributions with a loss in contrast at $\theta$
angles close to the poles.
%
\end{document}